\documentclass[11pt]{article}
\usepackage{amsfonts}
\usepackage{epsfig}

\newcommand{\person}[1]{{\sc #1}}

\newcommand{\R}{\ensuremath{\mathbb{R}}}
\newcommand{\N}{\ensuremath{\mathbb{N}}}

\newcommand{\normord}[2]{\,:\!\! {#1} \!\!:_{#2} \,}

\begin{document}

\begin{titlepage}
$\phantom{X}$\vspace{-20mm}\\ 
\noindent{$\phantom{X}$\hfill{\large\sc MS-TPI-99-13}}\vspace{20mm}
\begin{center}
{\bf\LARGE
$O(N)$-invariant Hierarchical
Renormalization Group Fixed Points by Algebraic
Numerical Computation and $\epsilon$-Expansion
}
\\[10mm]
{\large Johannes G\"{o}ttker-Schnetmann}\\[4mm]
{ 
September, 1999
\\[5mm]
Westf\"{a}lische Wilhelms-Universit\"{a}t M\"{u}nster\\
Institut f\"{u}r Theoretische Physik I\\
Wilhelm-Klemm-Str.~9\\
 D-48149 M\"{u}nster\\[1mm]
}
{\tt schnetm@uni-muenster.de}
\\[10mm]
{\bf Abstract} \\
\end{center}
Generalizing methods developed by {\sc Pinn},
{\sc Pordt} and {\sc Wieczerkowski} for the hierarchical
model with one component ($N=1$) and dimensions $d$ between $2$ and
$4$ we compute
$O(N)$-symmetric fixed points of the hierarchical renormalization
group equation for some $N$ and
$d$ with
$0 < d < 4$ and
$-2\le N \le 20$. The spectra of the linearized RG equation at the
fixed points are calculated and the critical exponents $\nu$ are
extracted from the spectrum and compared to Borel-Pad\'e-resummed
$\epsilon$-expansion. 
\\[4mm]
\end{titlepage}

\section{Introduction}

Euclidean quantum field theories and systems of
statistical mechanics are commonly defined by a generating
functional of correlation functions  
\begin{equation}
  {\mathcal Z}(J) = \int d\mu_v (\Phi) e^{-{\cal V}(\Phi)} e^{(J,\Phi)}
\end{equation}
where $d\mu_v$ is a suitable Gaussian measure with covariance
$v$ on the space of fields under consideration. The Gaussian measure
describes the free theory and $e^{-{\cal V}}$ describes the interaction and
will be called the ``Boltzmann factor''.

Let us consider a model with $N$ local degrees of freedom. Famous examples
of such models in statistical physics are the Ising model with $N=1$ and
the Heisenberg model with $N=3$. These models correspond to $\phi^4$
theories for fields with $N$ real components in the euclidean QFT setup.
Here we are using a local variable $\varphi \in\R^N$ to describe such
situations. 
We analytically continue our calculations to noninteger and 
negative values of $N$. The case $N=0$  is
an example of physical significance of this continuation, as it is used 
for the description of polymers.

$N$-component models have recently been reconsidered in the literature. In
this paper we are doing so using the high accuray scheme of the algebraic
renormalization group. Instead of fine tuning critical
parameters it is possible to calculate directly the fixed points by using
Newtons method. The critical exponents can then be calculated from the
spectrum of the linearized RG equation at the fixed point. We give a list
of scenarios for various values of
$N$ and $d$.

We will consider hierarchical
models in the sense of
\cite{pipowi,dipl} as an approximation to
``full'' field theoretic models models by replacing the
covariance $v = (-\Delta + m^2)^{-1}$ by a translationally
non-invariant hierarchical approximation of the massless
covariance. They can also be regarded as a system of
statistical physics with a peculiar structure. The
hierarchical covariance simulates the long distance behaviour
of the ``full'' covariance and preserves the locality of the
Bolzmannians under renormalization group transformations
(RGT). A Boltzmann factor is called local, if it factorizes
$$
   e^{-{\cal V}(\varphi)} = \prod_{x\in\Lambda}e^{-V(\varphi(x))} \;.
$$
The hierarchical RG equation is a RGT of Boltzmannians
$Z = e^{-V}:\R^N \rightarrow \R$. 

For $N$-component fields the hierarchical RG transformation
is given by the non-linear integral transformation 
\begin{equation}\label{rgequation}
 {\cal R}Z(\varphi) =  \int d\mu_{\gamma(1-\beta^2)}(\psi) Z^{L^d}(\psi +
\beta\varphi) \;,
\end{equation}
with the Gaussian measure $d\mu_\gamma(\varphi) = \frac
{1}{\sqrt{2\pi\gamma}^N} e^{-\frac {1}{2\gamma}\varphi^2}
d\varphi_1\ldots d\varphi_N$,
$\gamma > 0$, $\beta = L^{1-\frac d 2}$.
One may consider this equation as our starting point. A derivation of
equation (\ref{rgequation}) for $N=1$ can be found in \cite{pipowi}.
The derivation for general
$N$ can be done in an exactly analogous way.

In this paper we present the calculation of the
$O(N)$-symmetric fixed points and their spectrum. Fixed points of the RGT
determine the long distance behaviour of the system and are a tool to
investigate the continuum limit and the thermodynamic limit of such theories.
Recently also the stability of
$O(N)$-symmetric fixed points of this equation under
disturbances with cubic symmetry was investigated both for full
models and in the
hierarchical approximation
\cite{pirewi}. 


The critical
exponents are functions of $L$ in the hierarchical
approximation but they do not vary very rapidly with $L$
unless $L$ is close to 1. The limit $L\rightarrow 1$ is
perhaps singular. 
\person{Koch} and
\person{Wittwer} have shown the existence of a fixed point of
this equation with the values $N=1, d=3$ and $L^d=2$
\cite{kochwittwer}.  For
$L^d=2$ the model is equivalent to Dysons version of the hierarchical
model. Also a product of functions can be defined by
\begin{equation}
  (f\cdot g)(\psi) := 
\int d\mu_{\gamma(1-\beta^2)}(\psi) f(\psi +
\beta\varphi)
g(\psi +
\beta\varphi) \;.
\end{equation}
The hierarchical RG equation is then given  by
\begin{equation}
{\cal R}Z = Z\cdot Z  
\end{equation}
and the linearisation $L_{\cal R}$ of ${\cal R}$ at the ``point'' $f$ is
given by
\begin{equation}
  L_{\cal R}(f) Z = 2 f\cdot Z \;.
\end{equation}
Note that the product defined in this way is commutative but not associative 
\cite{nonassPordtWiec,dipl}. 

It is easy to see that there are
two trivial fixed points of the transformation ${\cal R}$ (apart from 
$Z_0 = 0$, which does not define a theory), namely
$Z_{UV} = 1$ and a Gaussian fixed point 
$$
Z_{HT}(\varphi) =
(2\beta^2)e^{\frac{2\beta^2-1}{4\gamma(1-\beta^2)}\varphi^2} \;.
$$
$Z_{UV}$ belongs to the massless free theory and $Z_{HT}$ to the massive free theory. One
important feature of the RG equation is
$$
  {\cal R_{\beta,\gamma}}(Z_{HT}Z) = Z_{HT} ({\cal R_{\beta',\gamma'}} Z)
$$
with $\beta'  = L^{-1-\frac d 2} = 2^{-\frac{2+d}{2d}}$ and $\gamma' = L^{-2} \gamma =
2^{-\frac 2 d}\gamma$. 
 The RG equation with
$\beta',
\gamma'$ is called the high temperature picture, because the HT fixed point is splitted
from the Boltzmannian. In the HT picture $Z=1$ belongs to the HT fixed point
and the inverse of
$Z_{HT}$ belongs to the UV fixed point. 

 The
linearisation at $Z_{UV}$ is given by
\begin{equation}
 \big(L_{\cal R}(Z_{UV}) f \big)(\psi)= 
\int d\mu_{\gamma(1-\beta^2)}(\psi) 1
f(\psi + \beta\varphi) \;.
\end{equation}
Normal ordering is in a sense inverse to
Gaussian integration. Using this one is able to compute the spectrum of
$L_{\cal R}(Z_{UV})$ analytically. This will be done in a first
section. After computing the spectrum we go on to discuss
``algebraic'' RG equations and their approximate solution using a
computer. Another way to compute fixed points is by
$\epsilon$-expansion. The
$\epsilon$-expansion is presumably not convergent. We use Borel-Pad\'e
summation of the first terms of the series upon the hypothesis that the
series is at least Borel summable. We conclude with a comparison of our
results with those found in the literature.

\section{The spectrum at $Z_{UV}$} \label{uvspectrum}

Since the massive theory is not critical the massless fixed point is more
interesting. The eigenfunctions at this fixed point are -- as in the case
$N=1$ -- Hermite (normal ordered) polynomials. Let $H_m$ be
the
$m$th Hermite polynomial given by
\begin{equation}
  H_m(\varphi) = (-1)^m e^{\frac{\varphi^2}{2}} \frac {d^m}{d\varphi^m}
e^{-\frac{\varphi^2}{2}}
= \sum_{j=0}^{\left[\frac m 2\right]}
\frac{(-1)^jm!}{(m-2j)!2^jj!}\varphi^{m-2j} = \normord{\varphi^m}{}
\end{equation}
and 
\begin{equation}
H_m^{(\gamma)}(\varphi) := \gamma^{m/2}H_m(\frac {\varphi}{\sqrt{\gamma}}) 
= \sum_{j=0}^{\left[\frac m 2 \right]}\frac{(-\gamma)^j
m!}{(m-2j)!2^jj!}\varphi^{m-2j}\;.
\end{equation}
Note that $H_m^{(0)}(\varphi) = \varphi^m$.

The $N$-component covariance is diagonal in color space. 
For a multiindex
$\mu = (m_1,\ldots,m_N)$ define
${\cal H}_\mu^{(\gamma)}$ by the $N$-fold tensor product of the
$H_{m_i}$
$$
{\cal H}_\mu^{(\gamma)}(\varphi_1,\ldots,\varphi_N) = 
H_{m_1}^{(\gamma)}(\varphi_1)\ldots H_{m_N}^{(\gamma)}(\varphi_N) 
= \normord{\varphi_1^{m_1}\ldots\varphi_N^{m_N}}{\gamma} \;.
$$
The normal ordered $O(N)$-invariant products $(\vec \varphi^2)^n$
are then given by
\begin{eqnarray*}
  \normord{(\vec\varphi^2)^n}{\gamma} & := & h_n^{(\gamma)}(\vec\varphi) :=
\sum_{\stackrel{\nu_1,\ldots,\nu_N}{\nu_1+\ldots+\nu_N = n}} {n \choose
\nu_1\ldots\nu_N} {\cal H}_{2\nu_1,\ldots,2\nu_N}^{(\gamma)} (\vec\varphi)\\
& = & \sum_{\stackrel{\nu_1,\ldots,\nu_N}{\nu_1+\ldots+\nu_N = n}} {n
\choose
\nu_1\ldots\nu_N}
\normord{\varphi_1^{2\nu_1}\ldots\varphi_N^{2\nu_N}}{\gamma}
\end{eqnarray*}
In the following we will use not use the notation
$\normord{(\vec\varphi^2)^n}{\gamma}$. Since normal ordering is the inverse of
Gaussian integration in the following sense 
$$
  \int d\mu_{\gamma(1-\beta^2)}(\psi) {\cal H}_\mu^{(\gamma)}(\psi +
\beta\varphi) = {\cal H}_\mu^{(\gamma-{\gamma(1-\beta^2)})}(\beta\varphi)
$$
and
$$
 {\cal H}_\mu^{(\beta^2\gamma)}(\beta\psi) = \beta^{|\mu|}{\cal
H}_\mu^{(\gamma)}(\psi)
$$
($|\mu| = m_1 + \ldots +m_N$) one finds that the $h_n^{(\gamma)}$ are a
complete set of eigenfunctions of $L_{\cal R}(Z_{UV})$ in the space
$L_2(\R^N, d\mu_\gamma)^{O(N)}$ of $O(N)$ symmetric functions. The
corresponding eigenvalues are $\lambda_n = 2\beta^{2n} =
2^{1+n\frac{2-d}{d}}$. The critical exponent
$\nu$ is calculated by
$$
  \nu = \frac{\ln(L)}{\ln(\lambda_1)} = \frac{\ln(2)}{d\ln(\lambda_1)} = 
 \frac 1 2 \;.
$$
We note that the spectrum in a space of $O(N)$ symmetric functions is
nondegenerate and the same as for the case $N=1$.

\section{Algebraic Computation}
\label{algcomp}

We now go on to the task of computing non-trivial fixed points and their
spectrum.
For calculations using a computer it is convenient to use the so called
algebraic RG equation. In this technical section we discuss the derivation
of this equation.

One expands the functions under considerations into a series of Hermite
polynomials or a power series as
$$
  Z(\varphi) = \sum_{n=0}^\infty z_n' h_n(\varphi) 
$$
or
$$
  Z(\varphi) = \sum_{n=0}^\infty z_n (\varphi^2)^n \;.
$$

These two possibilities to expand the functions minimize the necessary
work. Consider a general expansion
$$
  Z(\varphi) = \sum_{n=0}^\infty z_n e_n(\varphi) \;,
$$
where the $e_n$ are a suitable system of functions.
Then
\begin{eqnarray*}
  ({\cal R}Z)(\varphi) & = & 
\sum_{n,m=0}^\infty z_n z_m
\int d\mu_{\gamma(1-\beta^2)}(\psi)
e_n(\psi+\beta\varphi)e_m(\psi+\beta\varphi) \\
& = & 
\sum_{n,m,k=0}^\infty z_n z_m C_k^{nm}
\int d\mu_{\gamma(1-\beta^2)}(\psi)
e_k(\psi+\beta\varphi)
\,.
\end{eqnarray*}
For $e_n(\varphi) = (\varphi^2)^n$ the expansion of $e_n e_m$ is trivial and
 on the other hand if $e_n$ is a Hermite polynomial the integral is
trivial, since Gaussian integration is inverse to normal ordering.

\subsection{Expansion in powers of $\varphi^2$}
Let us consider first the expansion in powers of $\varphi^2$. One has
\begin{eqnarray}
  ({\cal R}Z)(\varphi) & = & \sum_{n,m} z_n z_m 
\int d\mu_{\gamma(1-\beta^2)}(\psi) \big((\psi + \beta\varphi)^2\big)^{n+m} \\
& = & \sum_{n,m} z_n z_m h_{n+m}^{(-\gamma(1-\beta^2))}(\beta\varphi) \label{expanded}
\end{eqnarray}
The next step is to express (\ref{expanded}) as a sum of
powers of
$\varphi^2$. To do the necessary calculations in a convenient way we
introduce the generating function $G^{(\gamma)}$ of the normal ordered
products.
$$
  G^{(\gamma)}(a,\varphi) := \sum_{n=0}^\infty \frac{a^n}{n!}
h_n^{(\gamma)}(\varphi)
$$
  By reordering of the series for $N = 1$ and induction to
general
$N$ one finds
$$
  G^{(\gamma)}(a,\varphi) 
 = \frac{1}{(1+2\gamma a)^{\frac N 2}}
\exp(\frac{a}{1+2\gamma a}\varphi^2) \;.
$$
Since $h_n^{(0)}(\psi) = (\psi^2)^n$ one expects $G^{(0)}(a,\varphi) = e^{a\varphi^2}$
which is easily verified. 


As a first step we want
to calculate the scalar products $\langle h_n^{(\gamma)},
h_m^{(\gamma)}\rangle_\gamma$ 
in $L_2(\R^N,
d\mu_\gamma)$. By expanding
$$
  \langle G(a,\cdot), G(b,\cdot)\rangle_\gamma = 
\int d\mu_\gamma(\varphi)
G(a,\varphi)G(b,\varphi) = (1-4\gamma^2ab)^{-N/2} 
$$
in powers of $a$ and $b$ and comparing coefficients, we find
\begin{equation}
\label{orthogonality}
  \langle h_n^{(\gamma)}, h_m^{(\gamma)} \rangle_\gamma = 
n!\prod_{i=0}^{n-1}(N+2i) (2\gamma^2)^n \delta_{n,m} \;.
\end{equation}
Equation (\ref{orthogonality}) can be generalized to
compute the coefficients of powers of
$\varphi^2$ in equation (\ref{expanded}). One calculates 
$$
  \langle G^{(\gamma')}(a,\cdot), G^{(\gamma)}(b,\cdot)\rangle_\gamma =
(1-2a(\gamma-\gamma')-4\gamma^2ab)^{-N/2} \;.
$$
By expansion into Binomial series two times and comparing coefficients we
find
$$
  \langle h_k^{(\gamma')}, h_n^{(\gamma)} \rangle_\gamma
= 
\left\{ \begin{array}{l@{\quad:\quad}l}
\prod_{i=0}^{k-1}(N+2i) n!
{k \choose n} (2\gamma^2)^n (\gamma-\gamma')^{k-n} 
  & n \le k\\ 0 & n > k \end{array}\right. \;.
$$
If we want to get the coefficients $c_n^{(k)}$ in 
\begin{equation}\label{hermiteexpansion}
  h_k^{(\gamma')} = \sum_{n=0}^\infty c_n^{(k)} h_n^{(\gamma)}
\end{equation}
as in equation $(\ref{expanded})$
we can use the last result together with the orthogonality of the $h_n^{(\gamma)}$
(\ref{orthogonality}) to find
$$
c_n^{(k)} = 
\left\{ \begin{array}{l@{\quad:\quad}l}
\prod_{i=n}^{k-1}(N+2i) 
{k \choose n} (\gamma-\gamma')^{k-n} 
  & n \le k\\ 0 & n > k \end{array}\right. \;.
$$
The derivation is restricted to $\gamma, \gamma' > 0$, but by
applying a Gaussian integration with covariance $\tilde\gamma$, the
covariances can be changed to $\gamma-\tilde\gamma$ and
$\gamma'-\tilde\gamma$.  We see that equation
(\ref{hermiteexpansion}) is correct for all values of $\gamma$ and
$\gamma'$. Inserting the coefficients $c_n^{(k)}$ into (\ref{expanded}) we
get
$$
  ({\cal R} Z)(\varphi) = \sum_{k=0}^\infty \sum_{n,m =
0}^\infty z_n z_m c_k^{(n+m)}
\beta^{2k} (\varphi^2)^k \;.
$$
By replacing $z_n$ by $z_n(\gamma(1-\beta^2))^{-1}$ that is using the expansion
$$
  Z(\varphi) = \sum_{n=0}^\infty \frac {z_n}{(\gamma(1-\beta^2))^n} (\varphi^2)^n
$$ we can
eliminate the Term
$(\gamma(1-\beta^2))^{n+m-k}$ from the final algebraic RG equation.
$$
  ({\cal R}Z)(\varphi) = 
     \sum_{k=0}^\infty
					\underbrace{
     \sum_{m,n}^\infty  
      z_n z_m
			   {\cal S}^{nm}_k(N)
      \beta^{2k}
      }
      _{({\cal R} Z)_k =: z'_k}
       (\gamma(1-\beta^2))^{-k}
      ((\varphi)^2)^k \;.
$$
$$
  {\cal S}^{nm}_k(N) := 
\left\{ \begin{array}{l@{\quad:\quad}l}
  \prod_{i=k}^{n+m-1}(N+2i) 
      {n+m \choose k} 
  & k \le n+m\\ 0 & k > n+m \end{array}\right. 
$$

The algebraic RG equation is a mapping of a subset of $\R^\infty$ to
$\R^\infty$ instead of an integral equation. It is given by
$$
  z = (z_0, z_1, z_2, \ldots) \mapsto {\cal R}(z)
$$
with
\begin{equation} \label{algrg}
  z_k \mapsto {\cal R}(z)_k = \beta^{2k}\sum_{n,m}^\infty {\cal
S}_k^{nm}(N) z_n z_m
\;.
\end{equation}
The dimension $d$ is only a numerical parameter in the integral
equation. But 
 equation (\ref{algrg}) contains
$N$ also only in numerical form. We can generalize to noninteger
values of $N$ and investigate other values of $N$ which are of
interest, e.g. $N=0$.  This is also possible for the integral equation
by integrating out the
$O(N)$-invariant part. The resulting equation is an integral equation with only one integration
containing a Bessel function and $N$ as a parameter.

\subsection{Solving the equation numerically}

To compute approximate solutions using a computer one simply
truncates the equation. We have to solve a finite system of
nonlinear equations of the type (\ref{algrg}). To avoid changing signs of the
coefficients in the expansion of the nontrivial fixed point we do the
calculations in the HT picture. Numerical
experience suggested as in \cite{pipowi} to use a different
normalization, namely
$$
  Z(\varphi) = \sum_{n=0}^\infty \frac{z_n}{2^{2n}\sqrt{(2n)!}}
(\gamma'(1-\beta'^2))^{-n} (\varphi^2)^n \;.
$$
The system of equations which was solved numerically was the one consisting of
$l_{max}+1$ equations ($k\in\{0,\ldots,l_{max}\}$) given by
$$
  z_k \mapsto z_k - {\beta'}^{2k} \sum_{n,m=0}^{l_{max}} S_k^{nm}(N) z_n z_m 
$$
and
{\footnotesize
$$
  S^{nm}_k(N) := 
\left\{ \begin{array}{l@{\quad:\quad}l} \prod\limits_{i=0}^{n+m-1}(N+2i)
      {n+m \choose k} 4^{k-n-m}\frac{\sqrt{(2k)!}}{\sqrt{(2n)!(2m)!}}
  & k \le n+m\\ 0 & k > n+m \end{array}\right. \;.
$$
}
This was done by the procedure {\sf C05PBF} of the NAG library, which is
a combination of the newton method and the gradient method to find zeros
of a set of nonlinear equations. The eigenvalues of the linearized RG
equation have then been computed by the procedure {\sf F02AFF}.
We first reproduced the results of \cite{pipowi} to check the correctness of the implementation.

\subsubsection{Effects of the truncation}
For $N=1, d=3$ an approximation of this type converges for $l_{max} \rightarrow \infty$
\cite{kochwittwer}. We investigated the effects of the truncation for general $N$ and there is
evidence that our approximation converges also. We found
it necessary to increase
$l_{max}$ with
$N$, see table
\ref{trunctable}.  For further discussion of the effects of the truncation see appendix
\ref{trunc}.

\begin{table}[ht]
{\small
\centerline{
  \begin{tabular}{|r||l|l|l|} \hline
  $l_{max}$ & $N = 3$ & $N = 10$ & $N = 20$ \\  \hline\hline
  $10$ & $0.77308443210437$ & - & - \\ \hline
  $20$ & $0.76113986112499$ & $0.92746045611772$ & - \\ \hline
  $30$ & $0.76113984902214$ & $0.91861185755319$ & $0.98467223300539$ \\ \hline
  $40$ & $0.76113984902214$ & $0.91861145661936$ & $0.96069946588199$ \\ \hline
  $50$ & -                  & $0.91861145661842$ & $0.96068060650581$ \\ \hline
  $60$ & -                  & -                  & $0.96068060512759$ \\ \hline
  $70$ & -                  & -                  & $0.96068060512757$ \\ \hline
  \end{tabular}
}
\caption{\small Effects of the truncation: $\nu$ as a function of $N$ and $l_{max}$.}
\label{trunctable}
}
\end{table}

\section{Numerical results}

We present here the observations made during the numerical investigation of the model. 
There is a bifurcation scenario of the hierarchical RG \cite{pipowi,
dipl, nonassPordtWiec}, which can be understood by means of the implicit
function theorem. 
At $d=4$ a double-well fixed point bifurcates from the UV fixed point,
which exists for $d < 4$. Generally at $d_k = \frac{2k}{k-1}, k\ge 2$
a
$k$-well fixed points bifurcates from the UV fixed point, which exists for
$d < d_k$. This can be understood more physically by observing that at $d_k$
the eigenvalue
$\lambda_k$ of the linearized RG equation at $Z_{UV}$ becomes relevant
($\lambda_k > 1$). This behaviour has indeed been  reproduced by our
numerical results for
$k=2$ and for some cases of $k=3$.

For plotting the fixed points one has to choose the parameter $\gamma'$. We
have used the same convention as \cite{pipowi}:
$$
  Z(\varphi) = \sum_{l=0}^{l_{max}} \frac {z_l}{2^{2l}\sqrt{(2l)!}}
(\gamma'(1-\beta'^2))^{-l}\varphi^{2l}
$$
with $\gamma' = \frac 1 2$. The plots show the potentials ($V = - \ln(Z)$) of
the fixed points in the UV picture normed by
$$
  Z(0) = 1 \;.
$$ 
$O(N)$-invariance allowes
us to use only one variable for integer $N$. We define fixed points
for noninteger $N$ by the formula above.

Close to the bifurcation points our truncation scheme becomes bad due to
the flatness of the non-trivial fixed point. We present here the
conjectured behaviour of the system, abstracted from the numerical results.
The fixed point which bifurcates at
$d=4$ from the UV fixed point has shown the following behaviour. 

\begin{figure}[p]
\centerline{
  \psfig{figure=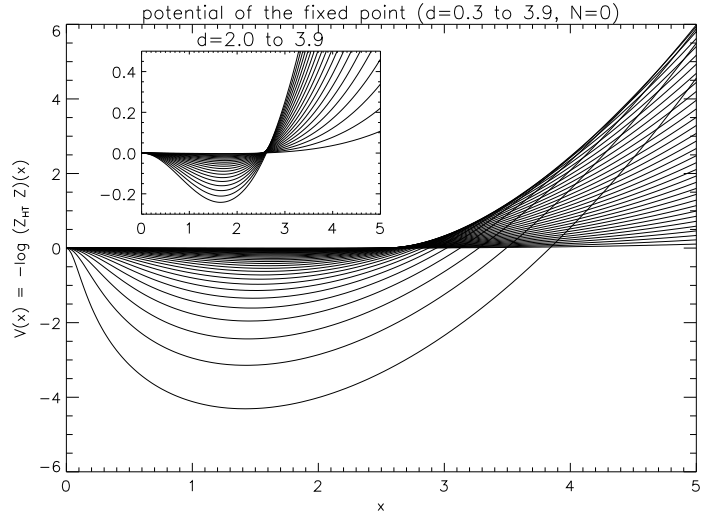,width=10cm}}
  \caption{\small Potentials of the fixed points ($N = 0, d = 0.3$ to
$3.9$, steps of
$\frac {1}{10}$). Deeper potentials belong to lower values of $d$. 
}
  \label{pot0}
\end{figure}

\begin{figure}[htp]
\centerline{
  \psfig{figure=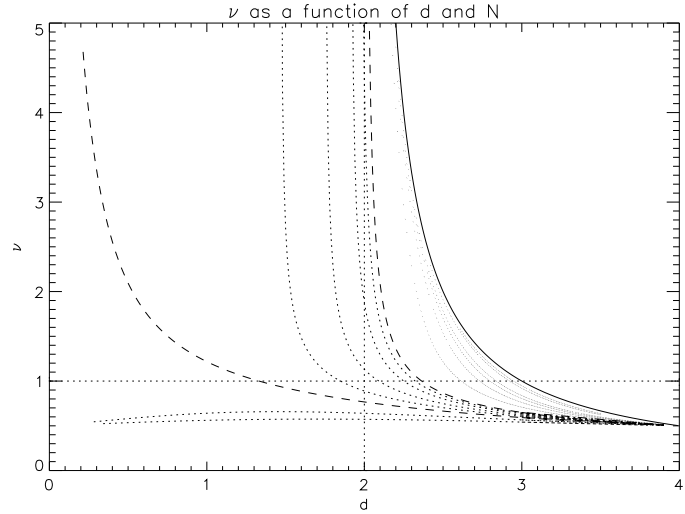,width=10cm}}
  \caption{\small The critical exponent $\nu$ for $0 < d < 4$ and (from the lower left to
the upper right)
$N = -1, -0.5, 0,
\frac 1 4, \frac 1 2, \frac 3 4, \frac 9 {10}$, $1$ to
$5$, $10$ and $15$. The plotted line 
$\frac{1}{d-2}$, is the asymtotic value of 
$\nu$ for $N \rightarrow\infty$  }
  \label{nuNd}
\end{figure}

\begin{figure}[htp]
\centerline{
  \psfig{figure=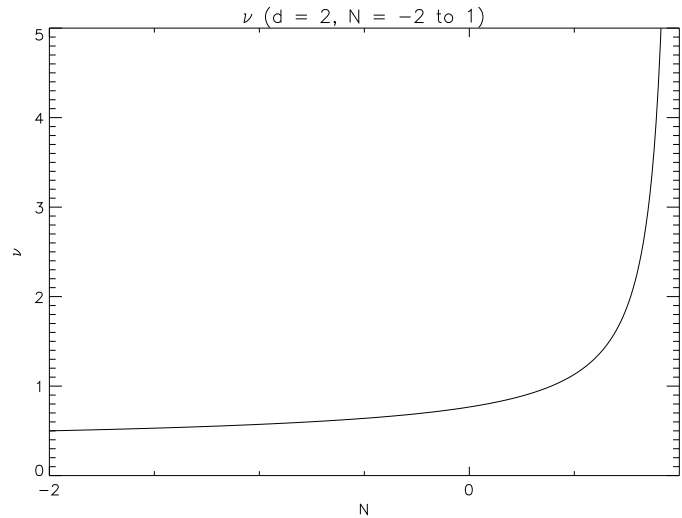,width=10cm}}
  \caption{\small The critical exponent $\nu$ for $d = 2$ and $N = -2$ to
$1$.}
  \label{nuNdiv2}
\end{figure}

\begin{figure}[htp]
\centerline{
  \psfig{figure=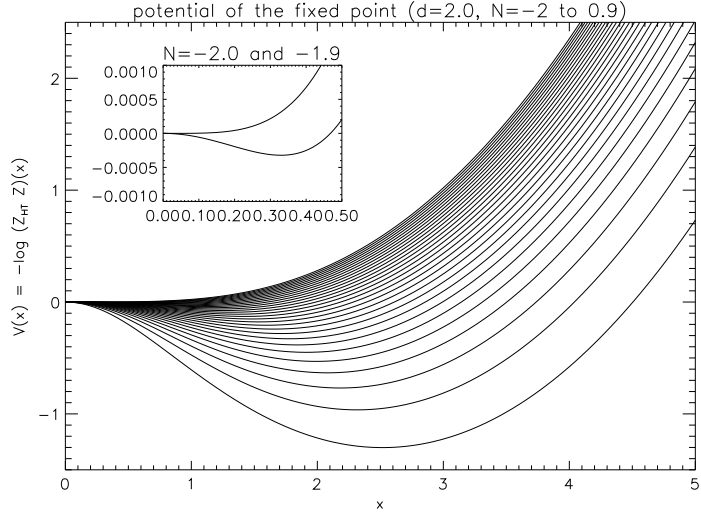,width=10cm}}
  \caption{\small Potentials of the fixed points ($d=2, N = -2$ to $0.9$, steps of $\frac
{1}{10}$). Deeper potentials belong to bigger values of $N$.}
  \label{nuNdiv2fix}
\end{figure}

\begin{figure}[htp]
\centerline{
  \psfig{figure=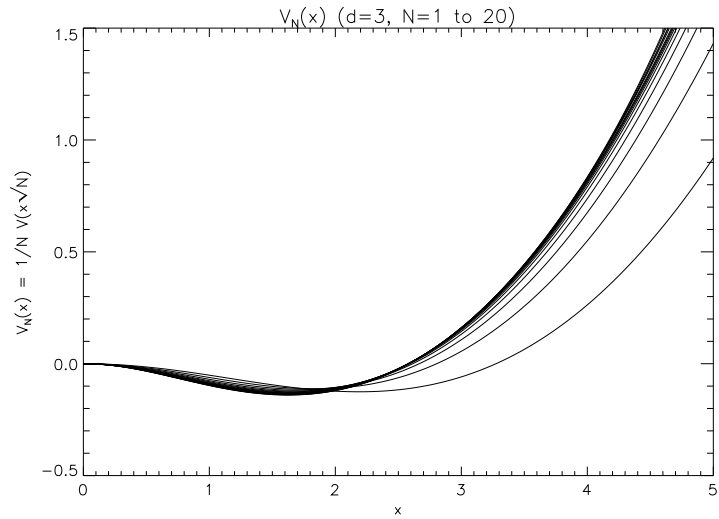,width=10cm}}
  \caption{\small Potentials of the fixed points ($d = 3, N=1$ to
$20$) after rescaling. The potential deviating most from the ``limit potential'' belongs
to $N=1$.}
  \label{pot30Scale}
\end{figure}

\begin{figure}[p]
\vspace{-1cm}
\begin{minipage}[t]{6cm}
\centerline{
  \epsfig{figure=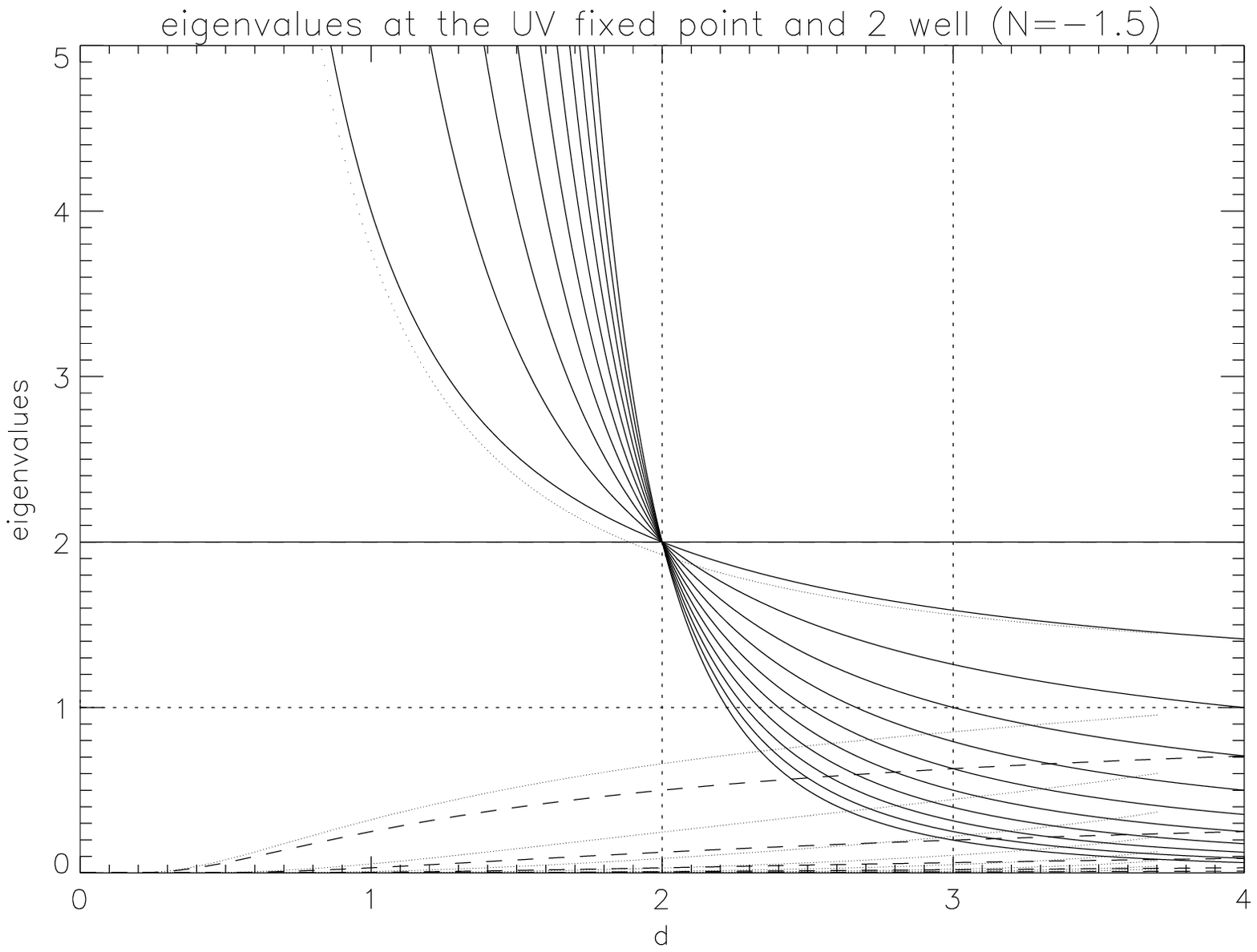,width=6cm}}
  \label{eigminus15}
\centerline{
  \epsfig{figure=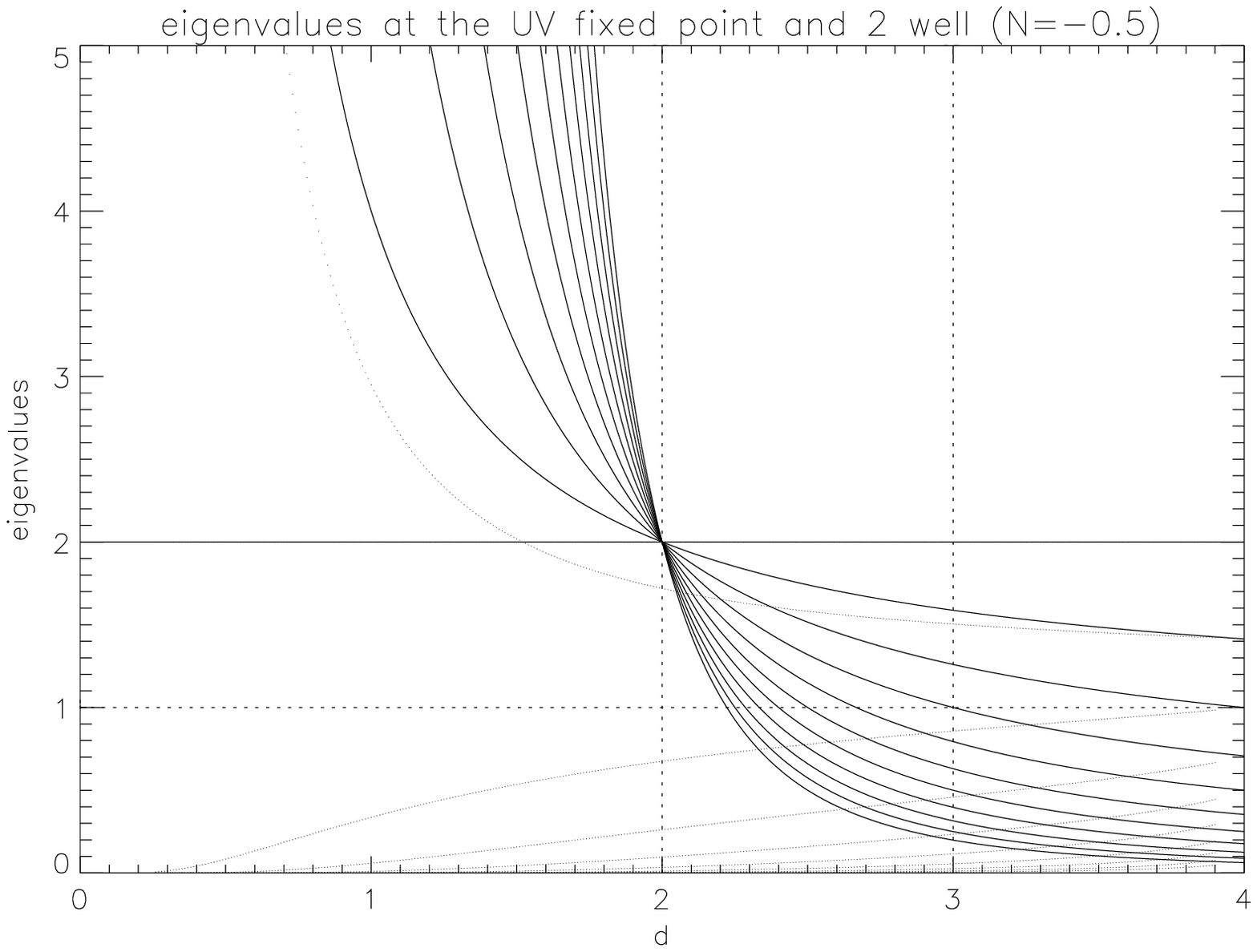,width=6cm}}
  \label{eigminus05}
\centerline{
  \epsfig{figure=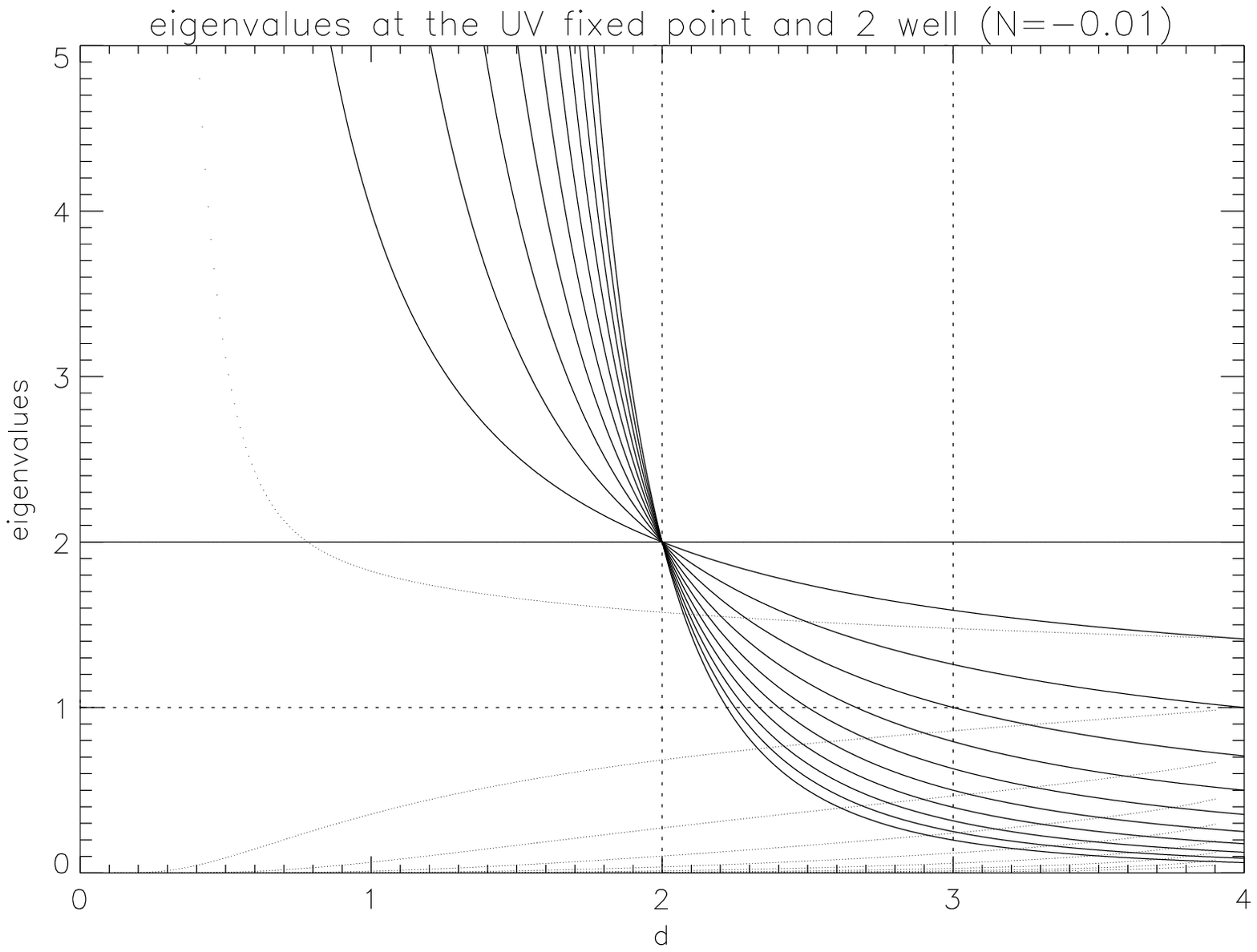,width=6cm}}
  \label{eigminus001}
\end{minipage}
\hfill
\begin{minipage}[t]{6cm}
\centerline{
  \epsfig{figure=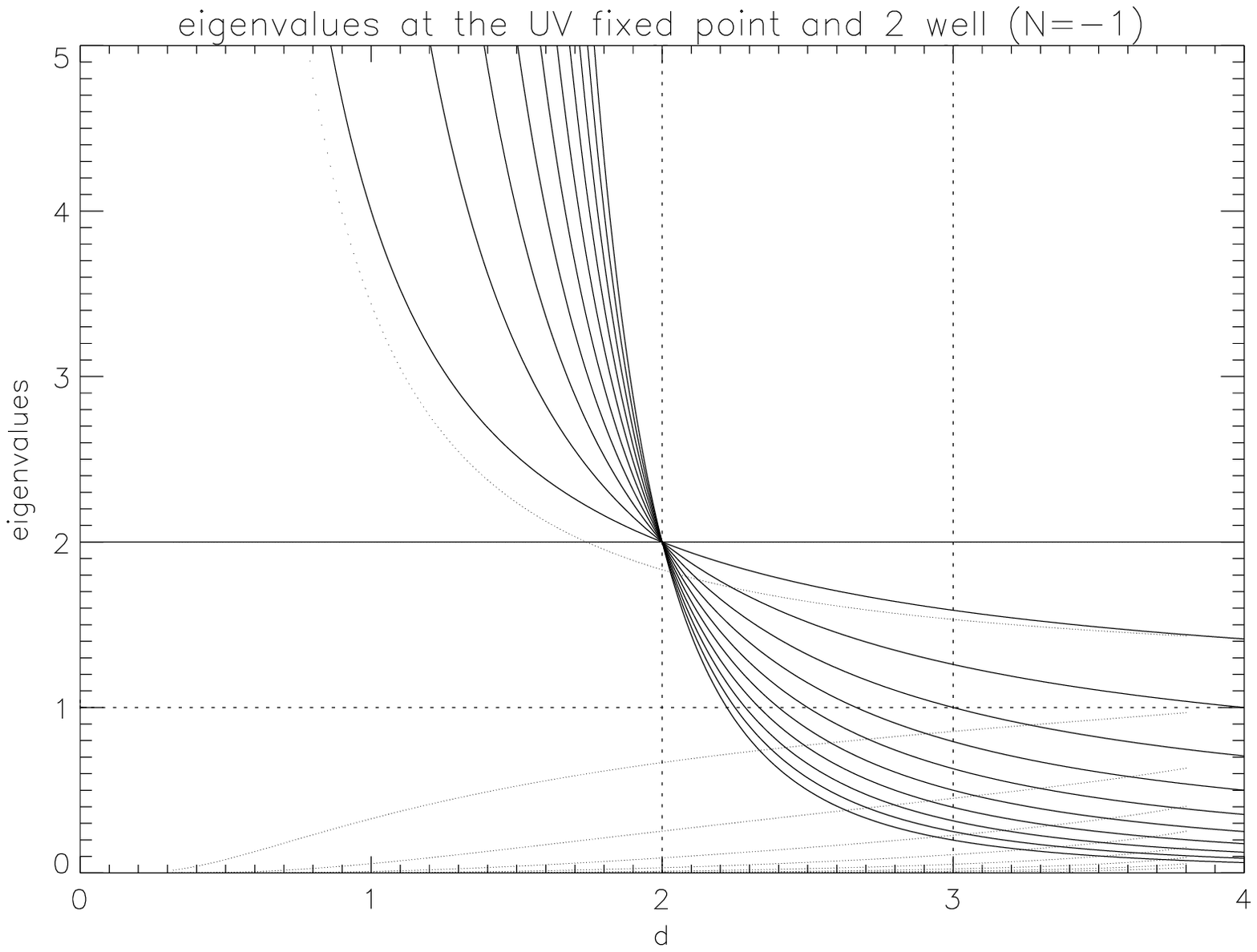,width=6cm}}
  \label{eigminus1}
\centerline{
  \epsfig{figure=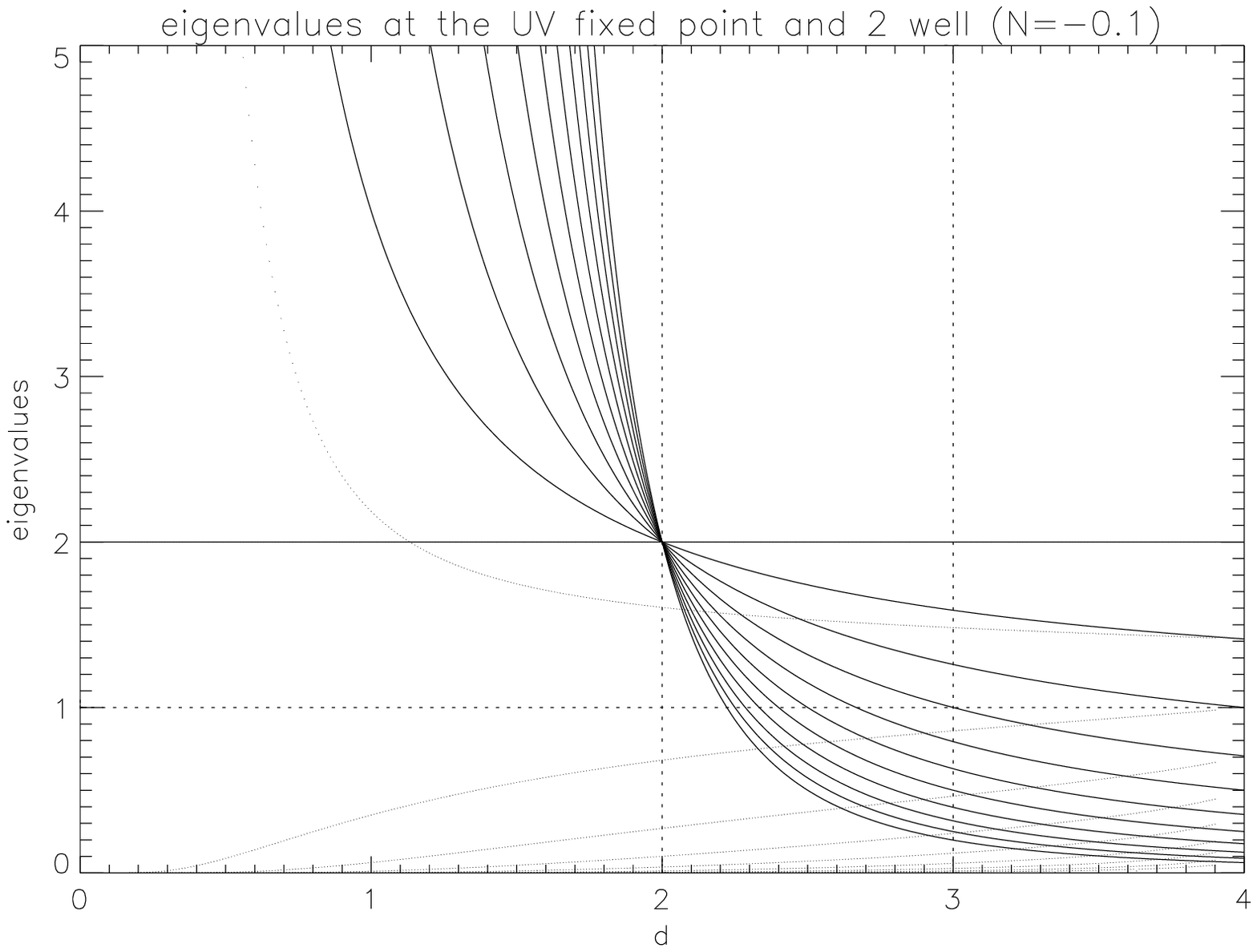,width=6cm}}
  \label{eigminus01}
\centerline{
  \epsfig{figure=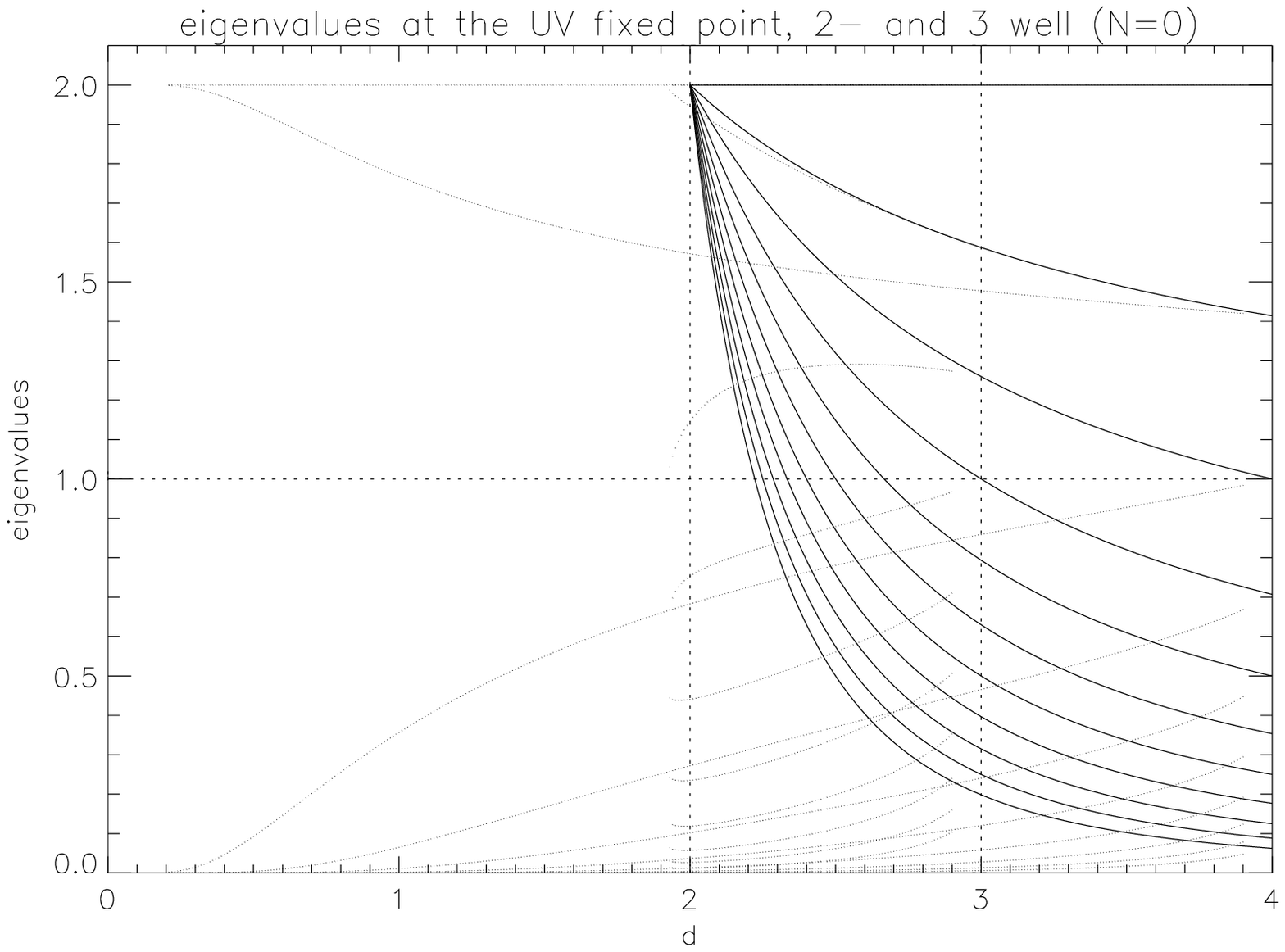,width=6cm}}
  \label{eig0}
\end{minipage}
\caption{\small Eigenvalues for $N \le 0$. The unbroken lines belong to
the eigenvalues at the UV fixed point, continued to $d < 2$. The 
($\cdots$) lines belong to the eigenvalues at the 2-well fixed point. The picture for 
$N=-1.5$ shows additionally  the eigenvalues at the HT fixed point (- -
-) and the picture for 
$N=0$ the eigenvalues of the 3-well fixed point ($\cdots$), which bifurcates at
$d=3$ from the UV fixed point. At every fixed point we also have 
a ``volume eigenvalue'' $\lambda_0 = 2$.}
\label{eigenkleiner0}
\end{figure}

\begin{figure}[ht]
\vspace{-1cm}
\begin{minipage}[t]{6cm}
\centerline{
  \epsfig{figure=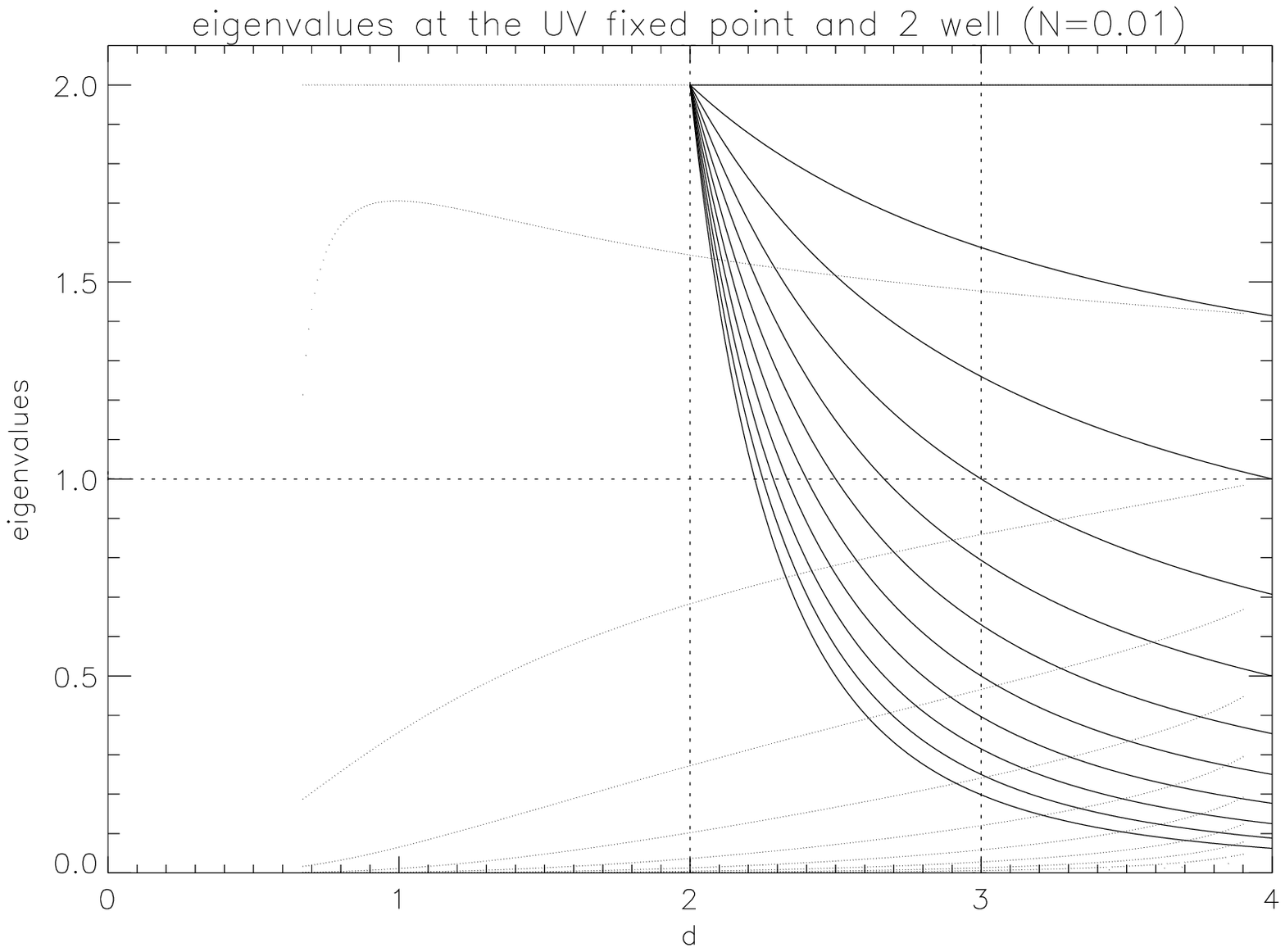,width=6cm}}
  \label{eig001}
\centerline{
  \epsfig{figure=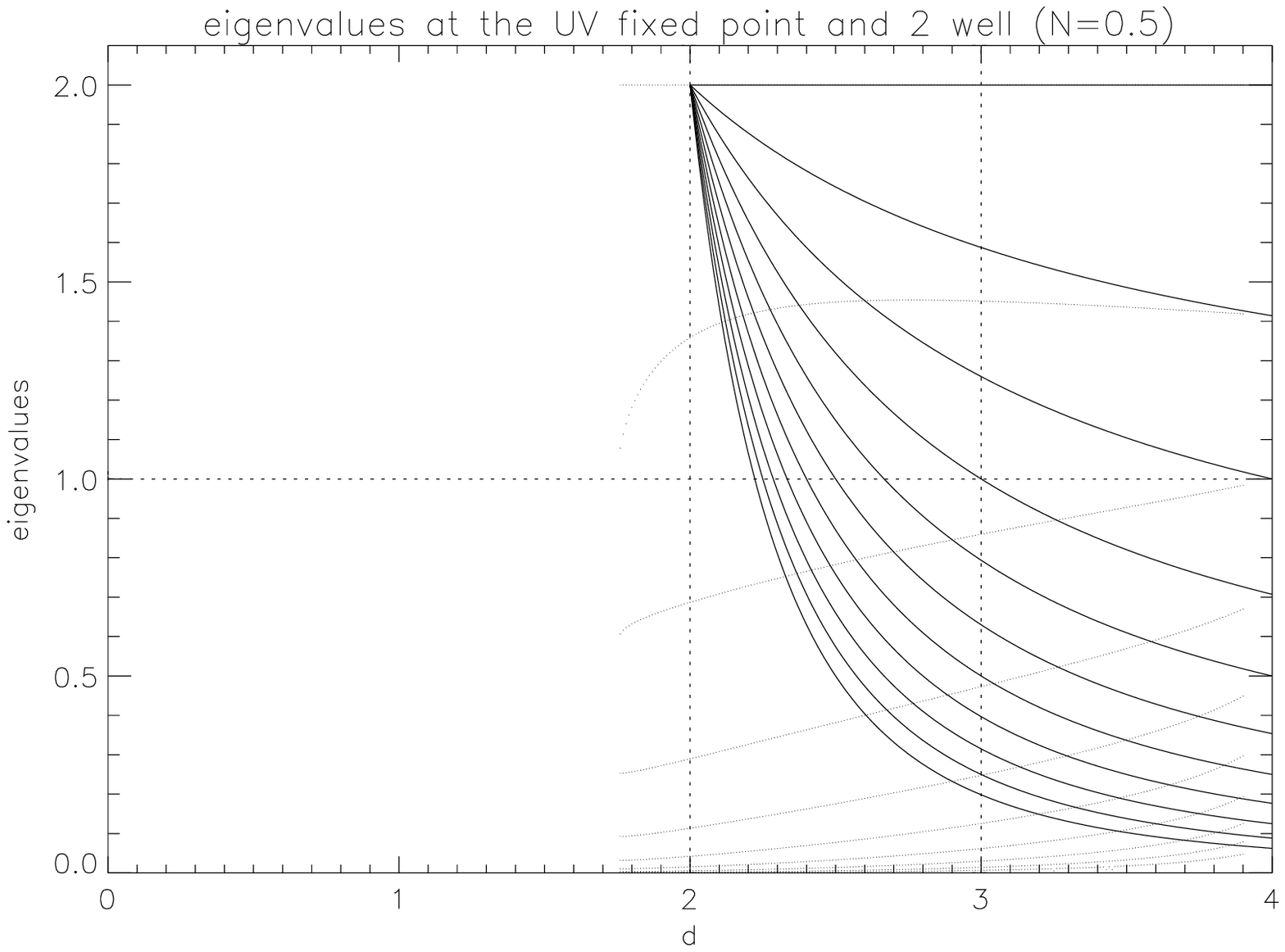,width=6cm}}
  \label{eig05}
\centerline{
  \epsfig{figure=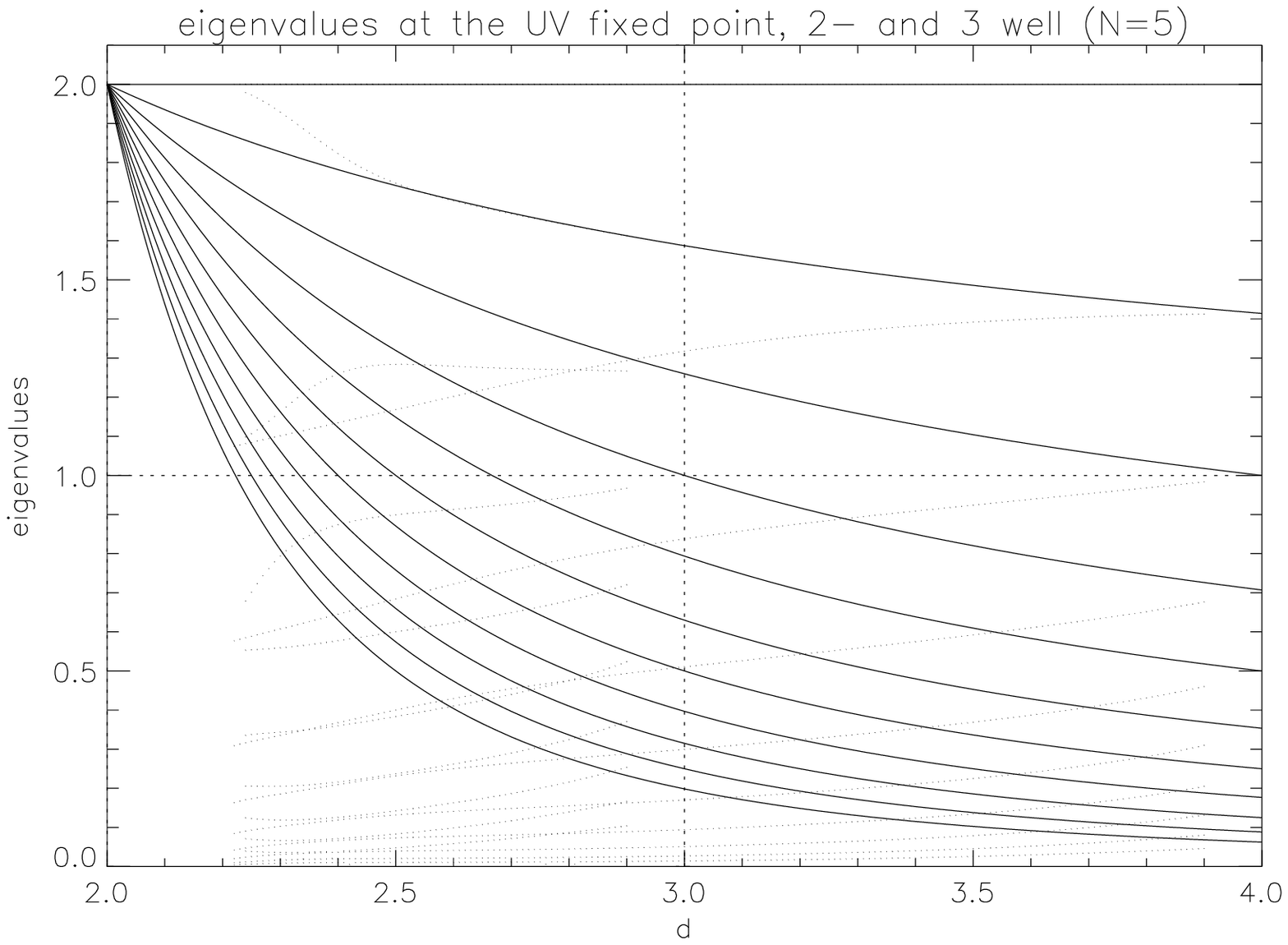,width=6cm}}
  \label{eig5}
\end{minipage}
\hfill
\begin{minipage}[t]{6cm}
\centerline{
  \epsfig{figure=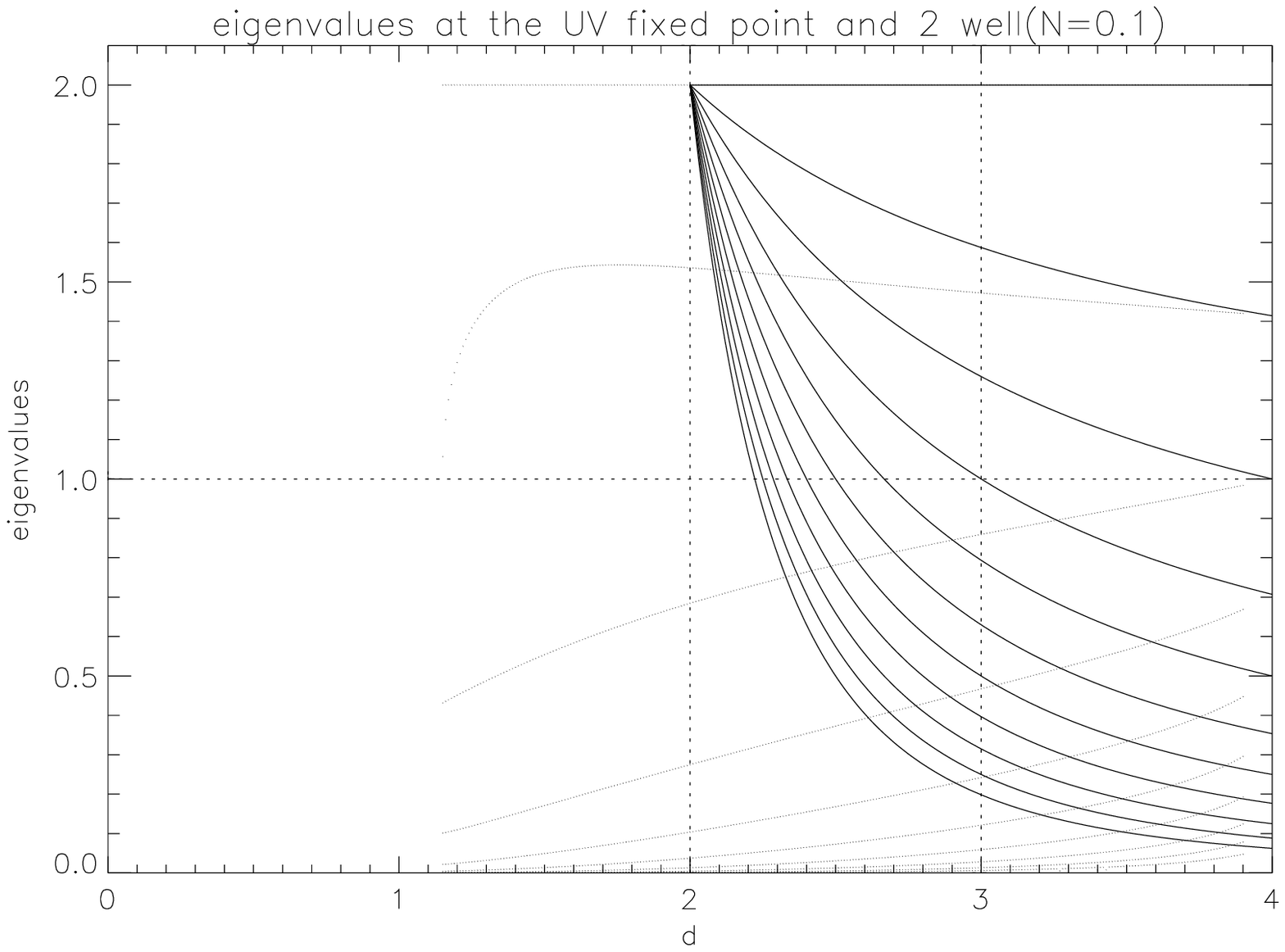,width=6cm}}
  \label{eig01}
\centerline{
  \epsfig{figure=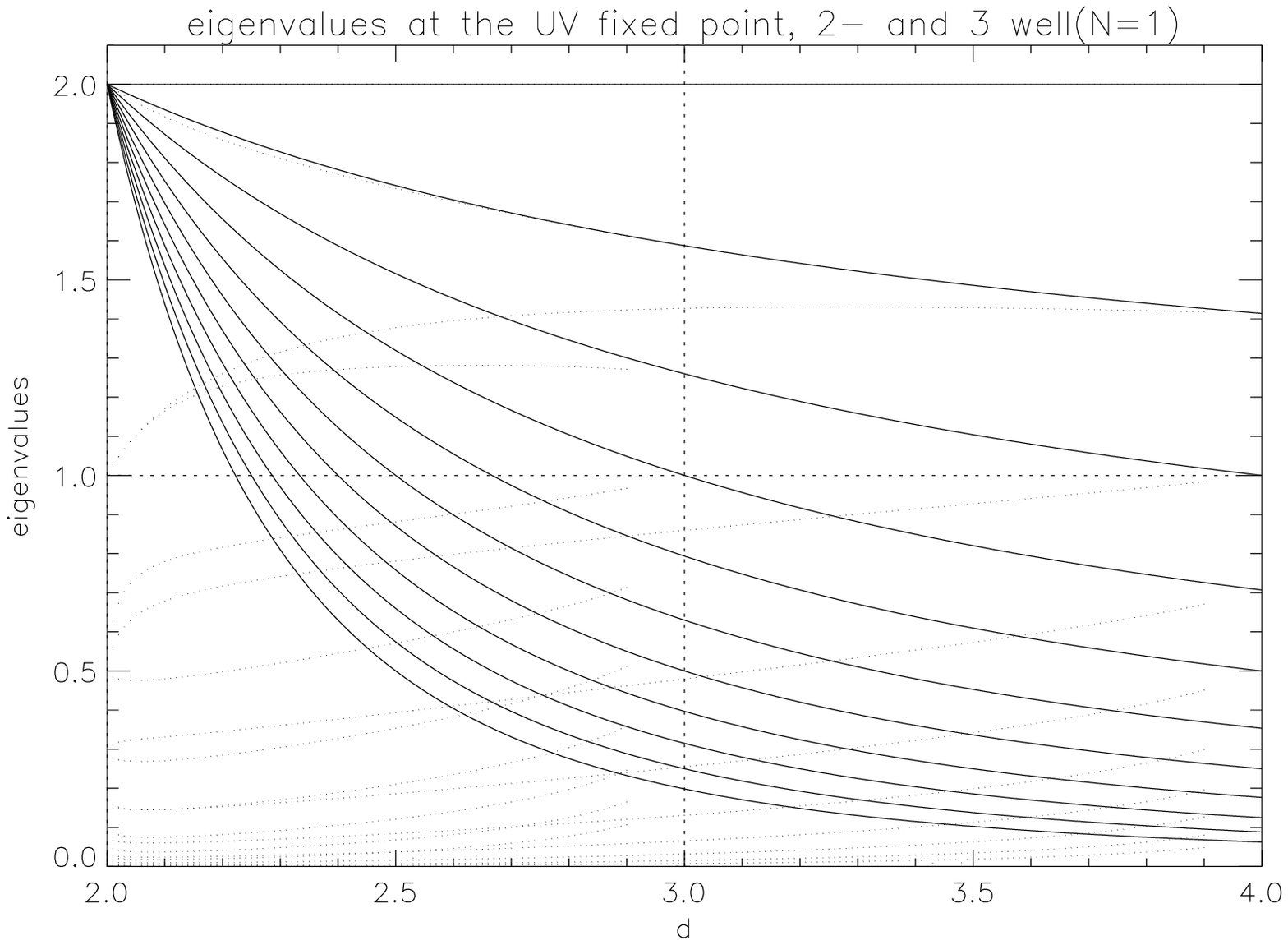,width=6cm}}
  \label{eig1}
\centerline{
  \epsfig{figure=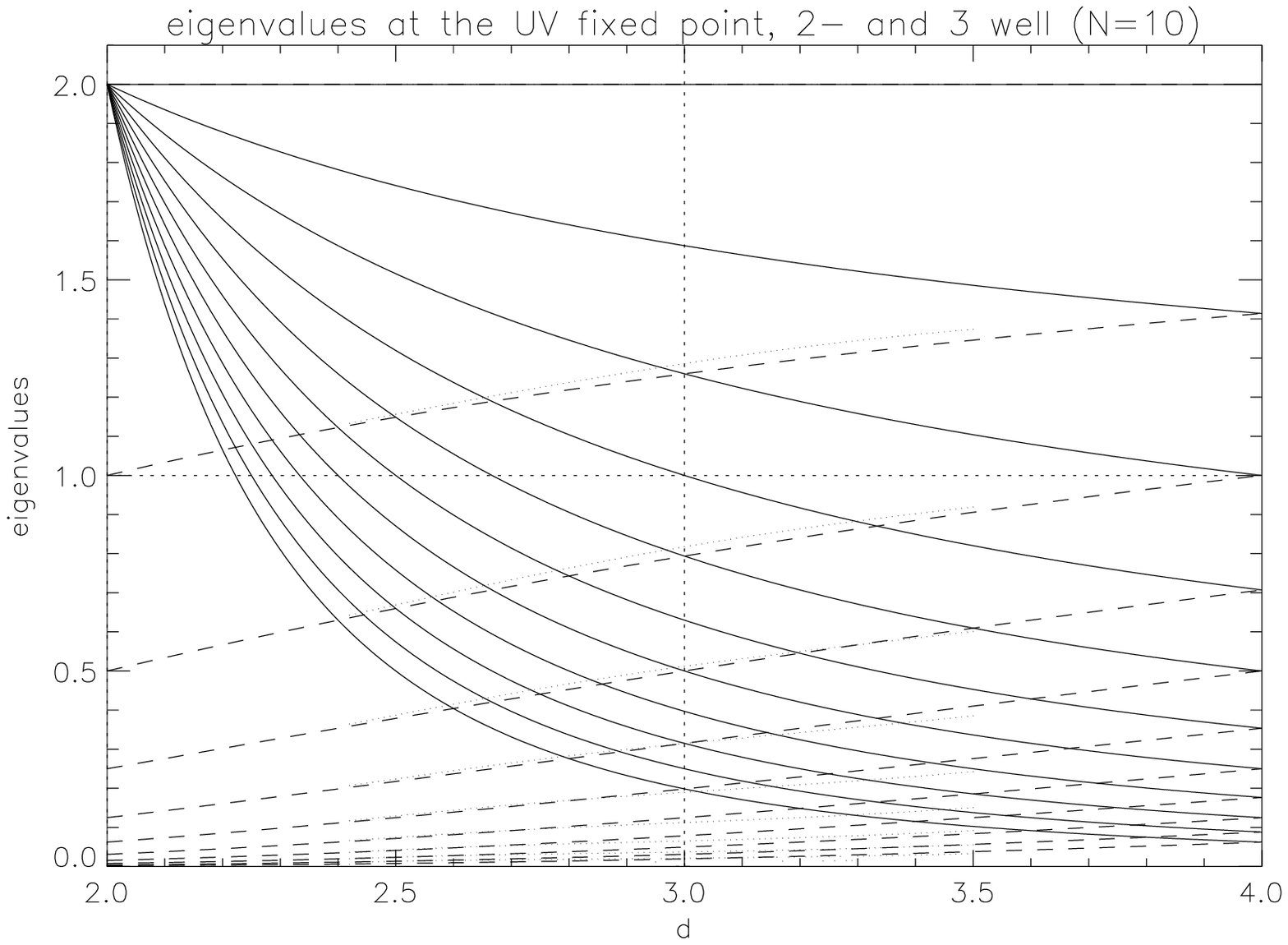,width=6cm}}
  \label{eig10}
\end{minipage}
\caption{\small 
Eigenvalues for $N > 0$. The unbroken lines belong to the
eigenvalues at the UV fixed point. The 
($\cdots$) lines belong to the eigenvalues at the 2-well fixed point. The picture for 
$N=10$ shows additionally  the eigenvalues at the $N\rightarrow\infty$ fixed point (- -
-) and the picture for 
$N=1$ and $N=5$ the eigenvalues of the 3-well fixed point ($\cdots$),
which bifurcates at
$d=3$ from the UV fixed point. At every fixed point we also have 
a ``volume eigenvalue'' $\lambda_0 = 2$.
}
\label{eigengroesser0}
\end{figure}

\begin{itemize}
\item For $N=-2$ 
the potential is a single well and the critical
exponent is $0.5$.  (Fig. 
\ref{nuNdiv2}, \ref{nuNdiv2fix}) 
The potential could be fitted to an approximatively quartic function
$\varphi^a$ ($a = 3.982(2)$) for $\varphi=0.01\ldots 0.5$ at $d=3$. 
First order $\epsilon$-expansion results in $v(\varphi) = c_1\;+c_2
\varphi^4$.

\item For $N>-2$ the potential is a double-well. The minimum becomes deeper with growing
$N$ and shifts to bigger arguments.
The critical exponent $\nu > \frac 1 2$ grows with $N$. 
$Z_*^{\frac 1 N}(\sqrt{N}\cdot)$ converges on a function $\zeta_*$ as $N\rightarrow\infty$.
(Fig. \ref{nuNd}, \ref{pot30Scale}
)

\item $\nu\rightarrow \frac 1 {d-2}$ as $N\rightarrow\infty$ and
$\nu\rightarrow\frac 1 2$ as $d\nearrow 4$ (Fig. \ref{nuNd}).

\item $N \ge 1$ (Fig. \ref{nuNd}, 
\ref{eigengroesser0})
  \begin{itemize}
    \item The fixed point exists for $2 < d < 4$.
    \item $\nu$ diverges as $d \searrow  2$, because the first fixed eigenvalue at the
fixed point converges on 1:
    $\lambda_1 \rightarrow 1$ in
    $
     \nu = \frac{\log 2}{\log(\lambda_1) d} \;.
    $
\end{itemize}

\item $0 < N < 1$ (Fig. \ref{nuNd}, \ref{eigengroesser0})
  \begin{itemize}
    \item The fixed point exists for $d' < d < 4$ with $0 < d' < 2$. 
    \item $d'$ grows (non linearly) with $N$ from $0$ to $2$. 
    \item $\nu$ diverges as $d \searrow d'$, because of $\lambda_1
\rightarrow 1$
    in $\nu = \frac{\log 2}{\log(\lambda_1) d}$. 
  \end{itemize}

\item $N = 0$. (Fig. \ref{pot0}, \ref{nuNd}, \ref{eigenkleiner0})
  \begin{itemize}
    \item The fixed point exists for $0 < d < 4$. 
    \item $\nu$ diverges as $d \searrow 0$, because of   
    $\frac 1 d \rightarrow \infty$ in $\nu = \frac{\log 2}{\log(\lambda_1) d}$. The first
eigenvalue converges 
$\lambda_1 \rightarrow 2$. 
  \end{itemize}
   
\item $-2 < N < 0$. (Fig. \ref{nuNd}, 
\ref{eigenkleiner0})
  \begin{itemize}
    \item The fixed point exists for $0 < d < 4$. 
    \item $\lambda_1$ diverges as  $d \searrow 0$. There is a
$0<d'<2$, with 
$\lambda_1 > \lambda_0 = 2$ for $0<d<d'$. $d'$ decreases (non linearly) with growing $N$
from $2$ to $0$.
    \item The limit of $\nu$ as $d \searrow 0$ exists, since the divergencies of
     $\lambda_1$ and $\frac 1 d$ cancel. 
  \end{itemize}

\item At $d=2$ $\nu$ diverges as $N\nearrow 1$, because of $\lambda_1
\rightarrow 1$. (Fig.
\ref{nuNdiv2},
\ref{nuNdiv2fix},
\ref{eigenkleiner0},
\ref{eigengroesser0}
).

\end{itemize}

With a view on the theorem of implicit functions $\lambda_1\rightarrow 1$ indicates
possible points of further bifurcations or the end of existence of solutions. We did
not find other solutions below the corresponding values of $d$ and is
was also not possible to follow the fixed point below these values of
$d$. Based on these observations the conclusion is, that these are the
points where the existence of the nontrivial fixed points ends. A
similar conclusion seems to apply for the case $d=2, N\nearrow 1$. The
fixed point which exists below $N=1$ does not exist for values of $N
\ge 1$. In this case other fixed points are known to exist at $N=1$,
$d=2$ \cite{nonassPordtWiec}.
\begin{table}[ht]
\centerline{
  \begin{tabular}{|c|c|c|c|c|c|} \hline
    $N = 0.01$ & $N = 0.1$ & $N = 0.25$ & $N = 0.5$ & $N = 0.75$ & $N = 0.9 $\\  \hline\hline
    $0.66(2)$ & $1.14(2)$ & $ 1.46(2)$ & $1.75(2)$ & $ 1.91(2)$ & $1.97(2)$\\ \hline
  \end{tabular}
}
\caption{\small Values $d'$ for which $\lambda_1(d') = 1$.}
\label{lambda1table}
\end{table}

Furthermore we investigated 3-well fixed points and found the following behaviour.

\begin{figure}[p]
\centerline{
  \psfig{figure=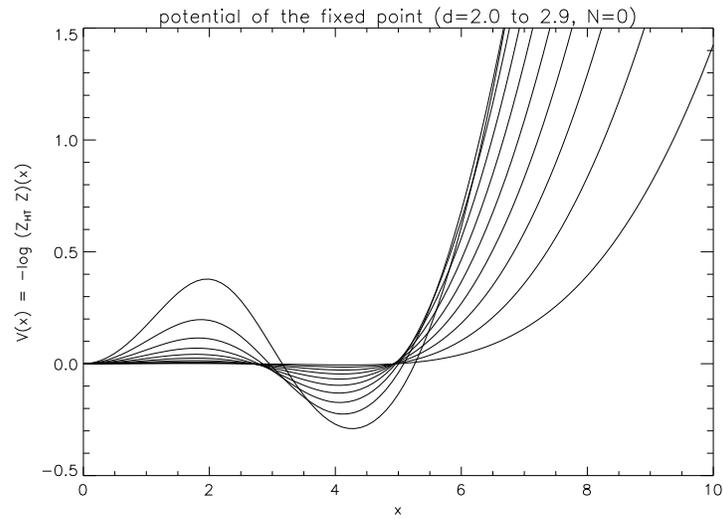,width=10cm}}
  \caption{\small Potentials of the 3-well fixed points ($N =
0, d=2.0$ to $d=2.9$). 
The local extremum grows with decreasing $d$.
}
  \label{three0}
\end{figure}

\begin{itemize}
\item There exists a 3-well fixed point for $N = 1$ and $2 <
d < 3$. (Fig. 
\ref{eigengroesser0} and the figures in \cite{pipowi})
\item There exists a 3-well fixed point for $N > 1$ and $d'(N) < d < 3$, $2\le d'(N) <
3$. It is unknown whether $d' > 2$. (Fig.
\ref{eigengroesser0})
\item There exists a 3-well fixed point for $N < 1$ and $d'
< d < 3$, $0 < d' < 2$. (Fig. \ref{three0}, \ref{eigenkleiner0})
\item The second eigenvalue $\lambda_2$ of the linearized RG equation at the
3-well converges on 
$1$ as $d\searrow d'$. (Fig. \ref{eigenkleiner0}, \ref{eigengroesser0})
\end{itemize}



\section{$\epsilon$-Expansion}

Following closely the lines of \cite{pipowi} one can generalize the
$\epsilon$-expansion to the model with $N$ components ($\epsilon = 4-d$). We work in the UV
picture because of the simple structure of the UV fixed point, 
$$
Z_{UV} = 1 = \sum_{k=0}^\infty z_n h_n^{(\gamma)} \;.
$$ 
This results in $z_n = \delta_{n,0}$. Furthermore we expand in terms of
hermite polynomials since they are the eigenfunctions at $Z_{UV}$, see
section
\ref{uvspectrum}. Again the first step is to calculate the algebraic RG
transformation for this case.
If we know the expansion of a product of two hermite polynomials we only have to integrate a
hermite polynomial, which is trivial. We have to calculate the coefficients 
${\cal C}_k^{nm}$ in
$$
  h_n^{(\gamma)} h_m^{(\gamma)} = \sum_{k=0}^\infty {\cal C}_k^{nm}h_k^{(\gamma)} \;.
$$
This is again done by the method of generating functions. Using the orthogonality equation
(\ref{orthogonality}) we get 
$$
  \langle h_n^{(\gamma)} h_m^{(\gamma)}, h_k^{(\gamma)} \rangle_\gamma
=  k! \prod_{i=0}^{k-1}(N+2i) (2\gamma^2)^k {\cal C}_k^{nm} \;.
$$
This leaves us with the calculation of the scalar product.
We calculate
$$
  \langle G(a,\cdot)G(b,\cdot),G(c,\cdot)\rangle_\gamma =
(1-4\gamma^2(ab+bc+ca)-16\gamma^3abc)^{-N/2}
$$
and do an a bit lengthy expansion of the right hand side of the equation to compare coefficients
of $a,b,c$ to find
\begin{eqnarray*} 
\lefteqn{\langle h_n^{(\gamma)} h_m^{(\gamma)},
                     h_k^{(\gamma)}\rangle_{\gamma}
            = n!m!k! (4\gamma)^{n+m+k} }\\
& &
   \sum_{\stackrel{n,m,k \le q}{2q \le n+m+k}} 
     \frac{\prod_{i=0}^{q-1}(N+2i)2^{-3q}}
          {(q-n)!(q-m)!(q-k)!(n+m+k-2q)!} \;.
\end{eqnarray*}
The final result is
\begin{eqnarray*}
\lefteqn{ {\cal C}_k^{nm}(N)  =    n!m! 4^{n+m+k} \gamma^{n+m-k} } & & 
   \\
   & &
   \sum_{\stackrel{n,m,k \le q}{2q \le n+m+k}} \prod_{i=k}^{q-1}(N+2i) 
  \frac{ 2^{-3q-k}}
          {(q-n)!(q-m)!(q-k)!(n+m+k-2q)!}
 \;.
\end{eqnarray*}
It was verified using {\sf\sl Maple} that for $N=1$ these coefficients are the same as those found
by \cite{pipowi}. The integration of $h_k^{(\gamma)}$ in the RG
equation is now trivial and only amounts to a multiplication with
$\beta^{2k}$. 

To make the algebraic RG equation again independent of $\gamma$ we use the
expansion
$$
  Z = \sum_{n=0}^\infty z_n \gamma^{-n} h_n^{(\gamma)} \;.
$$ 
Inserting this expansion into the RG equation results in
\begin{equation}
  {\cal R} Z  =
     \sum_{k=0}^\infty 
     \underbrace{\sum_{m,n=0}^\infty \beta^{2k}{\cal C}_k^{nm}(N) z_n z_m \gamma^{-n-m+k}}
     _{({\cal R} Z)_k =: z'_k} 
     \gamma^{-k}
     h_k^{(\gamma)}.
\end{equation}
We define the $\gamma$-independent structure coefficients $C_k^{nm}$ by
$$
  C_k^{nm} := {\cal C}_k^{nm}\gamma^{-n-m+k}
$$
and get the following algebraic RG equation 
$z = (z_0, z_1, \ldots) \mapsto {\cal R} z$ with
\begin{equation}
  z_k \mapsto ({\cal R}z)_k = \beta^{2k}\sum_{n,m = 0}^\infty C_k^{nm} z_n z_m \;.
\end{equation}

\subsection{$\epsilon$-expansion of the fixed points}

The $\epsilon$-expansion is an expansion of the coefficients $z_l$ of a fixed point into an
asymptotic series of the form
$$
  z_l = \sum_{k=0}^\infty z_l^{(k)}\epsilon^k \;.
$$
We write the fixed point equation in the form
$$
  \beta^{-2l}z_l = \sum_{n,m} C_k^{nm} z_n z_m \;.
$$
Expanding the left and the right hand side of the equation in powers of $\epsilon$ yields recursive
equations for the $z_l^{(k)}$. The calculation of these equations does not differ in an essential
way from the one in
\cite{pipowi}, and we will leave it out here. One only has to replace
the
structure coefficients by the $C_k^{nm}$ we just calculated. 

To first order we find for the fixed point
$$
  Z(\varphi) = 1-\frac{\ln 2}{16(N+8)}\gamma^{-2}\epsilon h_2^{(\gamma)}(\varphi) \;.
$$
For $h_2^{(\gamma)}$ we have
$$
  h_2^{(\gamma)}(\varphi) = \varphi^4 - 2\gamma(N+2)\varphi^2 + \gamma^2N(N+2)
\;.
$$
We note that for $N=-2$ and $\epsilon > 0$ the fixed point has the form of a
quartic single-well and for
$N>-2$ it is a double-well.

\begin{figure}[ht]
\begin{minipage}[t]{6cm}
\centerline{
  \epsfig{figure=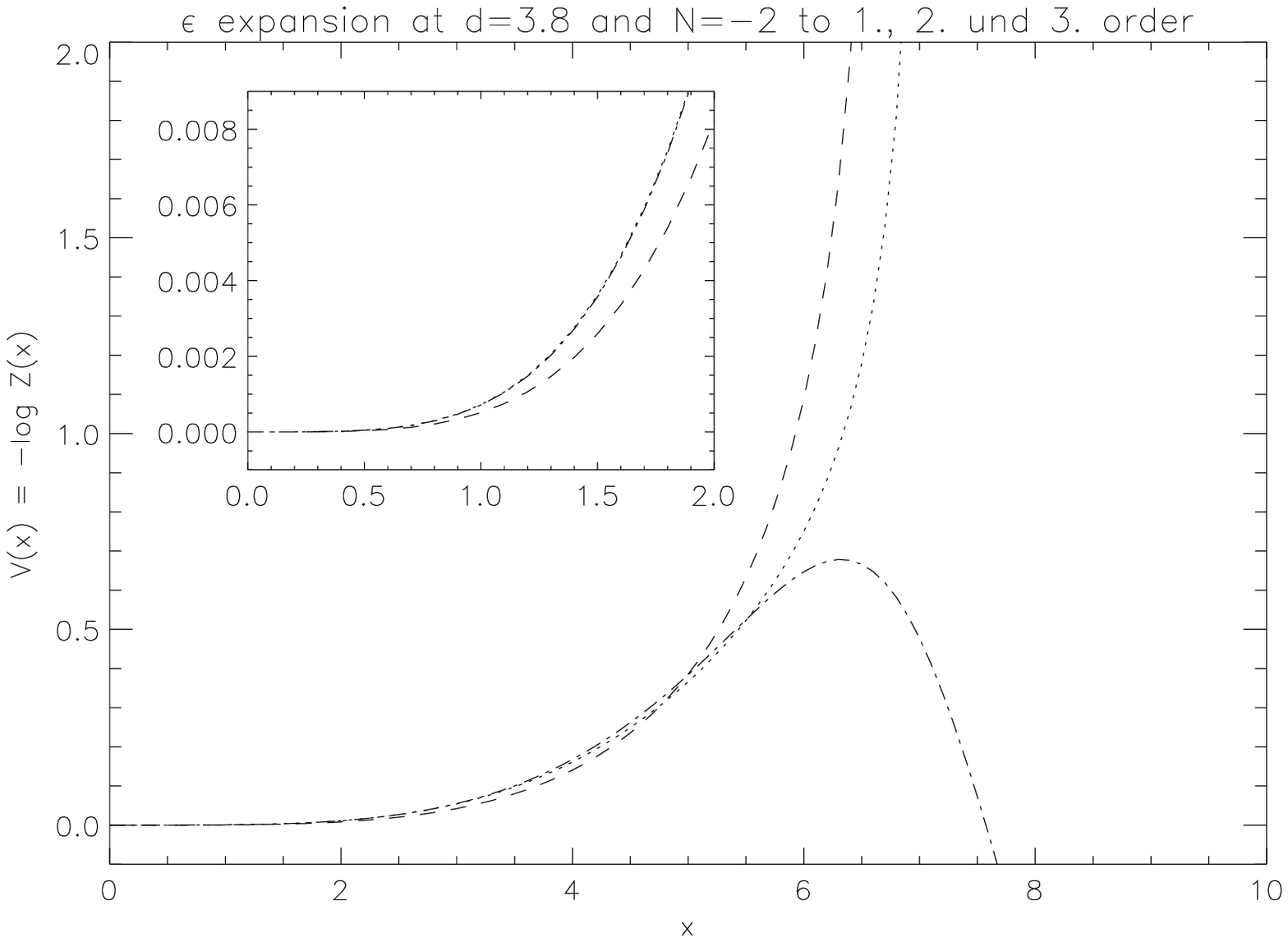,width=6cm}}
  \label{eps-2}
\centerline{
  \epsfig{figure=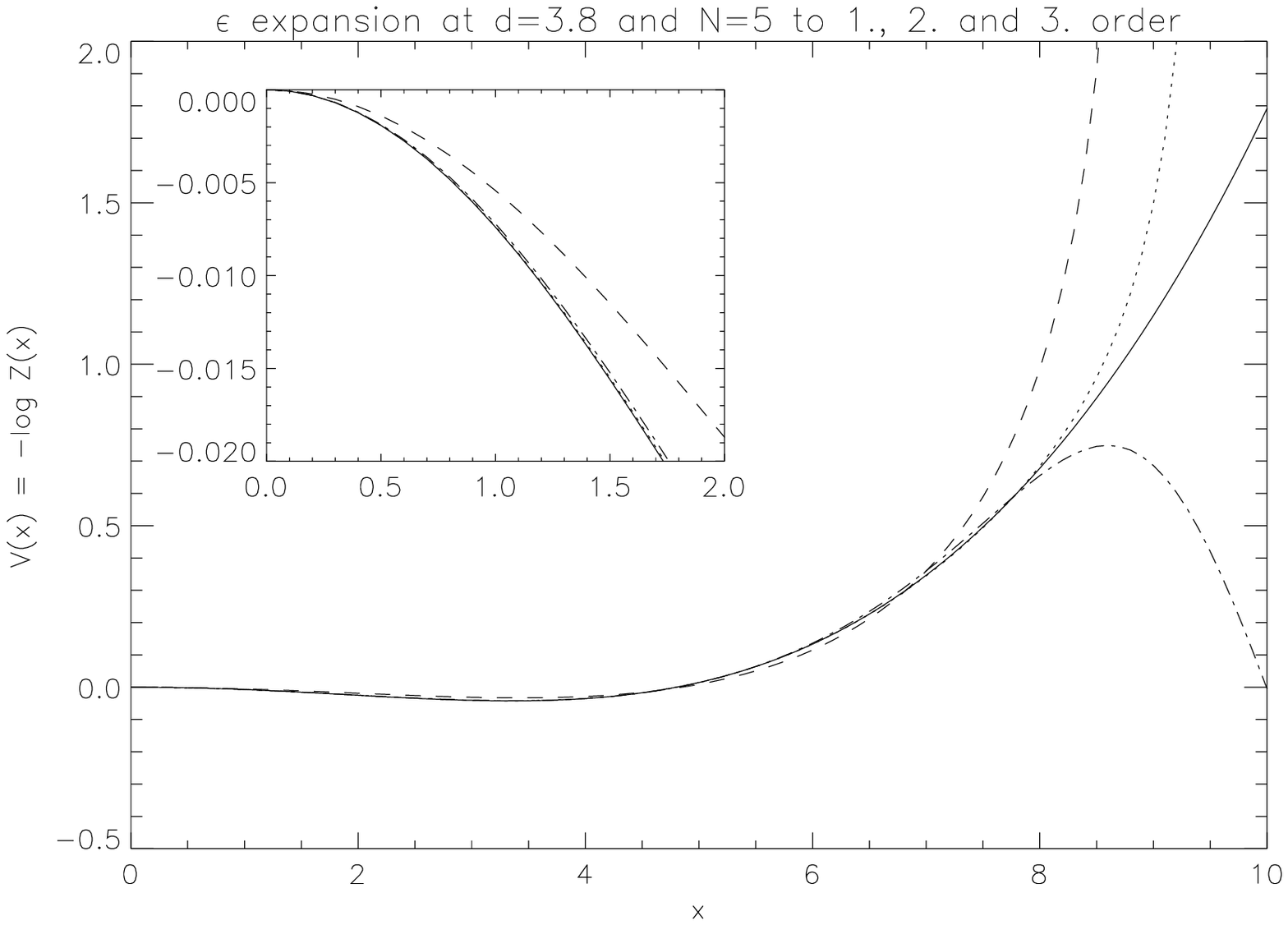,width=6cm}}
  \label{eps5}
\end{minipage}
\hfill
\begin{minipage}[t]{6cm}
\centerline{
  \epsfig{figure=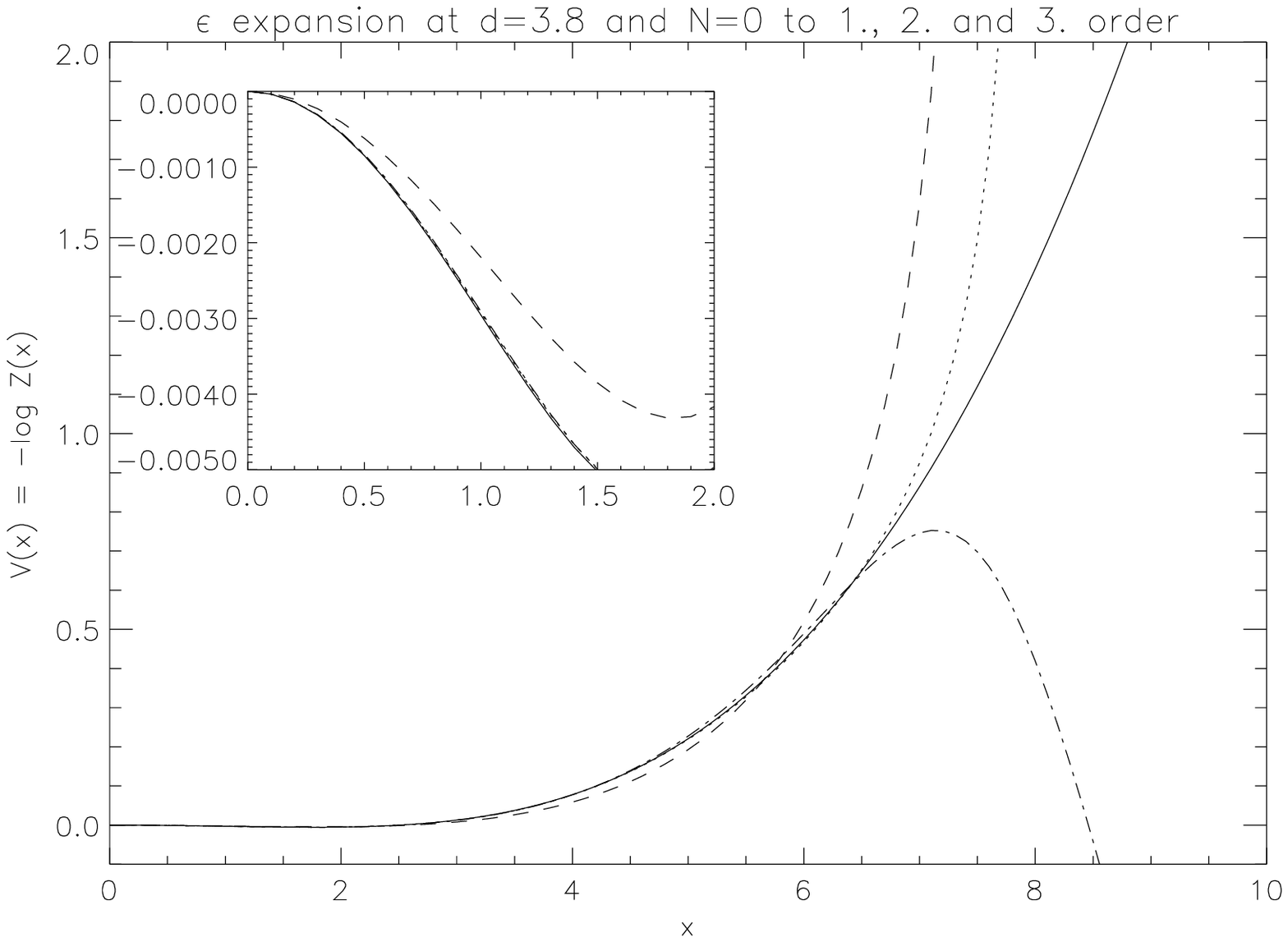,width=6cm}}
  \label{eps0}
\centerline{
  \epsfig{figure=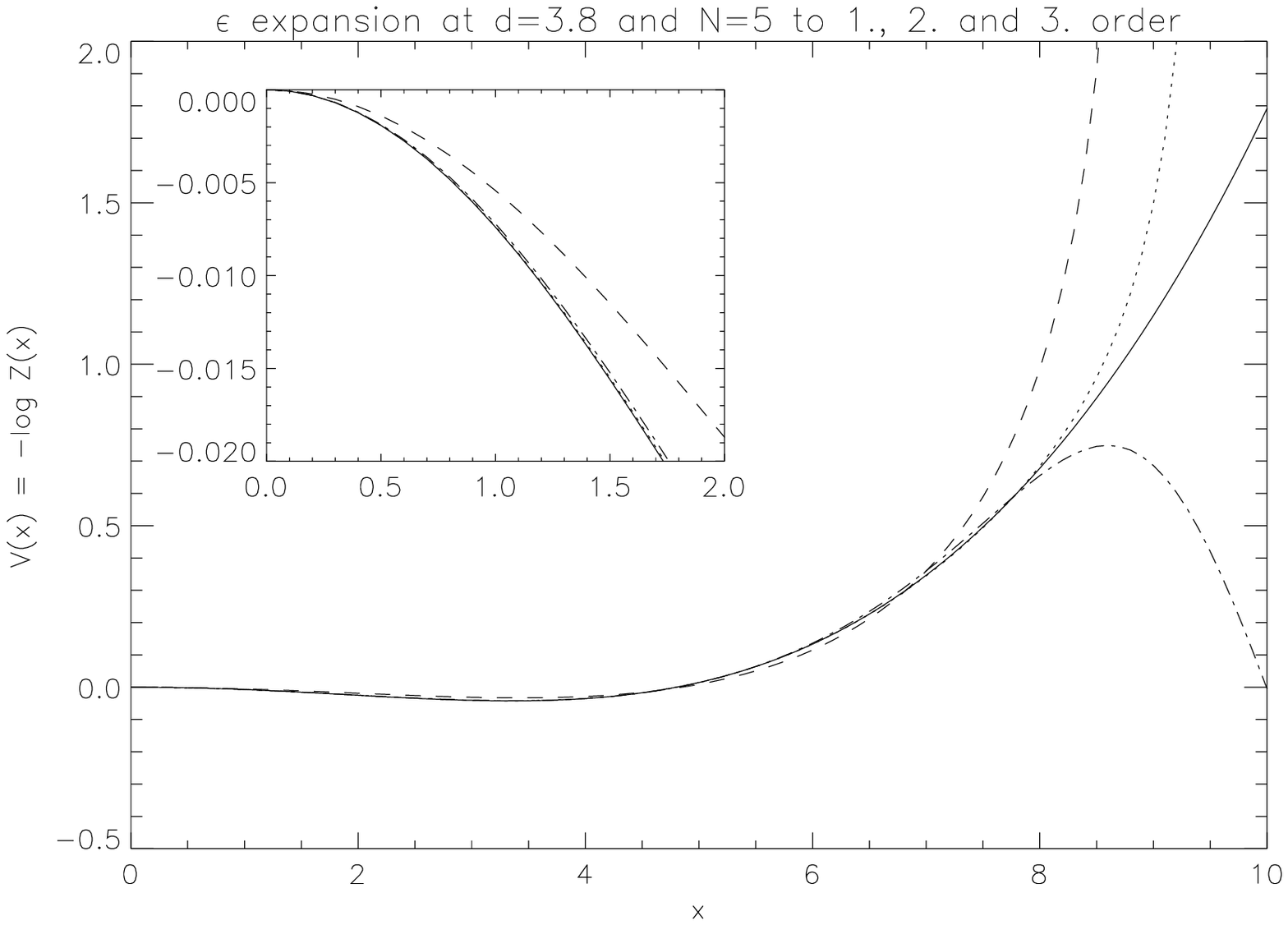,width=6cm}}
  \label{eps55}
\end{minipage}
\caption{\small $\epsilon$-expansion to 1. ($- - -$), 2. ($-\cdot-\cdot$) and 3. order
($\cdots$) compared to numerically calculated fixed points.}
\label{epsBilder}
\end{figure}

Figure \ref{epsBilder} presents some fixed points calculated to order 1, 2 and 3 by
$\epsilon$-expansion compared to numerically calculated fixed points.

\subsection{$\epsilon$-expansion of eigenvalues}

Given the fixed point we can calculate the linearisation of the RG equation at
this fixed point. The linearisation is given by the matrix ${\bf A}(z)$ 
$$ 
{\bf A}(z)_{ln} = 2\beta^{2l} \sum_{m=0}^\infty z_m C_l^{mn} \;.
$$
The eigenvalue equation is then
$$
  {\bf A}(z)v = \lambda v \;.
$$
We get the $\epsilon$-expansion of ${\bf A}(z)$ by inserting the expansions of $\beta^{2l}$
(with coefficients $b_{-l}^{(k)}$) and
$z_m$
\begin{equation}\label{matrixeps}
  {\bf A}(z)_{ln} = 2^{1-\frac{l}{2}}\sum_{k=0}^\infty \sum_{\nu=0}^k
    \sum_m C^{mn}_l(N)z_m^{(\nu)}b_{-l}^{(k-\nu)} \epsilon^k  \;.
\end{equation}
Expanding also the eigenvectors
$$
v_\mu  = 
\sum_{k=0}^\infty v_\mu^{(k)}\epsilon^k   
$$
and the corresponding eigenvalues
$$
\lambda_\mu  =  \sum_{k=0}^\infty \lambda_\mu^{(k)}\epsilon^k 
$$
we can use a generalisation of nondegenerate perturbation theory.
For details we refer to \cite{pipowi, dipl}. Summarizing we get the
following equations ($l_* = 2$)
\begin{equation} \label{AEntwickl}
  {\bf A}(z)_{ln}^{(k)} = 2^{1-\frac{l}{l_*}} 
    \sum_{\nu=0}^k \sum_m C^{mn}_l(N)z_m^{(\nu)}b_{-l}^{(k-\nu)} 
\end{equation}
\begin{equation} \label{nullteordnung}
  v_\mu^{(0)} = (\delta_{n,\mu}) \quad\mbox{and}\quad \lambda_\mu^{(0)} =
2^{1-\frac{\mu}{l_*}}
  \;,\;\; \mu \in \N_0
\end{equation}
\begin{equation} \label{eigenvektorentwickl2}
0 \stackrel{!}{=} (v_\mu^{(0)}, v_\mu^{(k)}) =
\sum_{n=0}^\infty \delta_{\mu,n} v_{\mu, n}^{(k)} = v_{\mu, \mu}^{(k)}
\end{equation}
\begin{equation} \label{lambdaEntwickl}
 \lambda_\mu^{(k)}  = 
   \sum_{n=0}^{k-1} \sum_{m=0}^\infty v_{\mu,m}^{(0)} 
   \sum_{j=0}^\infty{\bf A}^{(k-n)}_{mj} v_{\mu,j}^{(n)}
   = 
   \sum_{n=0}^{k-1} 
   \sum_{j=0}^\infty{\bf A}^{(k-n)}_{\mu j} v_{\mu,j}^{(n)} \;,
\end{equation}
\begin{equation} \label{eigenvektorentwickl}
   v_{\mu, \rho}^{(k)} = 
   \frac {1}{2(2^{-\frac{\mu}{l_*}} - 2^{-\frac{\rho}{l_*}})}
   \sum_{n=0}^{k-1} \Bigg( \sum_{m=0}^\infty  
   \Big( {\bf A}^{(k-n)}_{\rho m} v_{\mu, m}^{(n)} \Big)  - 
  \lambda_\mu^{(k-n)} v_{\mu, \rho}^{(n)}  \Bigg)
\end{equation}
With these equations it is possible to calculate the $\epsilon$-expansion of the eigenvalues
and eigenvectors recursively order by order. To each order the sums are finite.

The critical exponent $\nu$ belongs to $\lambda_1$. The first order coefficient is 
$$
  \lambda_1^{(1)}  =  \sqrt{2} \frac{\ln 2}{4} 
    \Big(\frac{1}{2} - \frac{N+2}{N+8} \Big) \;.
$$
Summing up to first order 
$$
  \lambda_1 = \sqrt{2}
    \Bigg(1 + \frac{\ln 2}{4}
    \Big(\frac{1}{2} - \frac{N+2}{N+8} \Big) \epsilon \Bigg) \;,
$$
and expanding the logarithm in $\nu^{-1} = \frac {(d_* - \epsilon) \ln
\lambda_1}{\ln 2}$ we get
$$
  \nu^{-1} = 2\Bigg(1 - \frac 1 2 \frac{N+2}{N+8} \epsilon \Bigg) + O(\epsilon^2) \;,
$$
so that 
$$
 \nu = \frac 1 2 + \frac 1 4 \frac{N+2}{N+8} \epsilon + O(\epsilon^2) \;.
$$

The $\epsilon$-expansion was calculated up to 6th order using {\sf Maple V
Release 3}. The results of \cite{pipowi} for the case $N=1$ were reproduced. 
In appendix \ref{epsilonappendix} some analytical results of the $\epsilon$-expansion are
given.

\begin{table}[htp]
{\footnotesize
\centerline{
  \begin{tabular}{|r||c|c|c|c|c|c|c|c|} \hline
   $d$   & $k=1$ & $k=2$ & $k=3$ & $k=4$ & $k=5$ & numerical & BP \\ 
\hline\hline
$3.9$ & $ 0.5065 $ & $ 0.5066 $ & $ 0.5066 $ &
$ 0.5066 $ & $ 0.5066 $ & $ 0.50659 $ & $ 0.50659 $
\\
$3.8$ & $ 0.5135 $ & $ 0.5140 $ & $ 0.5137 $ &
$ 0.5139 $ & $ 0.5137 $ & $ 0.51380 $ & $ 0.51380 $
\\
$3.7$ & $ 0.5211 $ & $ 0.5222 $ & $ 0.5211 $ &
$ 0.5221 $ & $ 0.5208 $ & $ 0.52157 $ & $ 0.52157 $
\\
$3.6$ & $ 0.5293 $ & $ 0.5313 $ & $ 0.5286 $ &
$ 0.5319 $ & $ 0.5261 $ & $ 0.52990 $ & $ 0.52988 $
\\
$3.5$ & $ 0.5382 $ & $ 0.5412 $ & $ 0.5359 $ &
$ 0.5442 $ & $ 0.5266 $ & $ 0.53877 $ & $ 0.53873 $
\\
$3.4$ & $ 0.5477 $ & $ 0.5521 $ & $ 0.5430 $ &
$ 0.5602 $ & $ 0.5176 $ & $ 0.54820 $ & $ 0.54811 $
\\
$3.3$ &  $ 0.5580 $ & $ 0.5641 $ & $ 0.5496 $ &
$ 0.5821 $ & $ 0.4945 $ & $ 0.55821 $ & $ 0.55761 $
\\
$3.2$ & $ 0.5691 $ & $ 0.5771 $ & $ 0.5555 $ &
$ 0.6128 $ & $ 0.4557 $ & $ 0.56884 $ & $ 0.56841 $
\\
$3.1$ & $ 0.5810 $ & $ 0.5913 $ & $ 0.5607 $ &
$ 0.6570 $ & $ 0.4045 $ & $ 0.58012 $ & $ 0.57946 $
\\
$3.0$ & $ 0.5940 $ & $ 0.6068 $ & $ 0.5651 $ &
$ 0.7236 $ & $ 0.3488 $ & $ 0.59209 $ & $ 0.59040 $ 
\\ \hline
  \end{tabular}
}
}
\caption{ $\nu(N=0)$ calculated by different methods explained in the
text.}
\label{epsbpkaptable}
\end{table}

We calculated $\nu$ first in the ``naive'' way, that is by summing the
$\epsilon$-expansion of
$\lambda$ and calculating $\nu=\frac{\ln 2}{d \ln\lambda_1}$. 
This is called ``naive'', because the $\epsilon$-expansion is presumably not
convergent. Upon the hypothesis that it is at least Borel-summable, we
calculated the Borel sum of the expansion
$$
  B(\epsilon) = \sum_{k=0}^{\mbox{\tiny max. order}} \frac
{\lambda_1^{(k)}}{k!} \epsilon^k \;.
$$
Then we calculated the 
diagonal Pad\'{e} approximation
$Q(\epsilon)$ of this sum and did a numerical Borel transform of $Q$
$$
  \lambda_1(\epsilon) = \int_0^\infty Q(t \epsilon) \exp(-t) dt \;.
$$

Table \ref{epsbpkaptable}  shows values of the critical exponent $\nu$ calculated by ``naive''
summation of the $\epsilon$-expansion up to order
$k$, by numerical computation of the eigenvalues and by Borel-Pad\'{e} summation of the
$\epsilon$-expansion up to 
 6th order. 

Table \ref{epsbptable} shows values of $\nu$ calculated by Borel-Pad\'{e}
approximation of the 
$\epsilon$-ex\-pansion up to 6. order. 
For small $\epsilon$ there is a very good agreement between the values calculated by solving the
truncated system and those calculated by Borel-Pad\'{e} summation of the
$\epsilon$-expansion.  If we go to 
$d=3$ only 2 to 3 digits agree.  Unfortunately we do not have error estimates for the
Borel-Pad\'{e} summation.


\section{Comparison with other results and Conclusion}

\begin{table}[htp]
{\footnotesize
\centerline{
  \begin{tabular}{|l|llllll|} \hline
   & $N = -1$ & $N = 0$ & $N = 1$ & $N = 2$ & $N = 3$ & $N=4$ \\  \hline\hline
   & $0.541536$ & $0.592086$ & $0.649570$ & $0.708225$ & $0.761140$ & $0.804364$ \\ \hline
  \cite{pipowi} 
     & - & - & $0.64957$ & - & - & - \\ \hline
  \cite{comellastravesset} Pol
     & - & - & $0.6496$ & $0.7082$ & $0.7611$ & $0.8043$ \\ \hline
  \cite{comellastravesset} WH
     & - & $0.6066$ & $0.6895$ & $0.7678$ & $0.8259$ & $0.8648$ \\ \hline
  \cite{wegnerhoughtonrg} WH
     & - & - & $0.6896$ & $0.767$ & $0.826$ & $0.865$ \\ \hline
\hline
  \cite{reiszhightemperatureON} HT
     & - & - & $0.6301(18)$ & $0.6734(28)$ & $0.7131(40)$ & $0.7361(68)$ \\ \hline
  \cite{guidazinnjustin} pert. 
     & - & $0.5882(11)$ & $0.6304(13)$ & $0.6703(13)$ & $0.7073(30)$ & $0.7410(60)$ \\
\hline
  \end{tabular}
}
}
\caption{\small Comparison of values of $\nu$ for $d=3$.}
\label{litexpvgl}
\end{table}

Table 3 of \cite{comellastravesset} contains values of the critical
exponent
$\nu$ calculated from the Wegner-Houghton and the Polchinski RG
equations. It is remarkable that our values coincide to three significant
digits with those found by the Polchinski RG equation. See table
\ref{litexpvgl}. 
The results of the Wegner-Houghton
equation differ from our results, but are compatible with those in \cite{wegnerhoughtonrg}.
Compared to results for the ``full'' model the hierarchical model
overestimates the critical exponent.  We refer to the results of
\cite{reiszhightemperatureON} found by convergent high temperature expansion and of
  \cite{guidazinnjustin} calculated by perturbation theory. 
For $N\rightarrow\infty$ and $N\rightarrow -2$ the
hierarchical model has the same exponents as the ``full'' model.


We have presented the calculation of $O(N)$ symmetric fixed points of the hierarchical RG
transformation by $\epsilon$-expansion and by numerical methods based on
the expansion of functions. 
The results of the $\epsilon$-expansion are compatible with those found by numerical
computation when $\epsilon$ is not to large. Furthermore the results are compatible with
results found by investigating the Polchinski equation.

Considering this, $O(N)$ symmetric solutions of the hierarchical approximation seem to be
quite well understood. 
Mathematically there is still room for
improvement since rigorous proofs of the existence of the fixed points calculated in this
paper are to my knowlege still
lacking for most values of $N$ and $d$ with the exception $N=1, d=3, L^d = 2$. 
\person{Cassandro} and
\person{Mitter} have found other fixed points for $d <
2$ and $N\ge 1$ \cite{RGApproachToSurfacesCassandro}. It should be possible to find these
fixed points by algebraic computation. In this work no fixed points for these values of $N$
and $d$ have been calculated. It would be interesting to know if at least for
$2 < d
\le 4$ the bifurcation scenario leads to all fixed points of the RG equation. 

Other - more physical - routes from this point are systematic
corrections to the hierarchical approximation 
and the study of fixed points with less
symmetry, e.g. \cite{pirewi}.

\section{Acknowledgements}

I would like to thank Dr. A. Pordt and Dr. C. Wieczerkowski for helpful
discussions on hierarchical models and Dr. C. Wieczerkowsi for
reading the manuscript.

\begin{appendix}
\section{Effects of the truncation}
\label{trunc}

\begin{table}[htp]
\centerline{
{\tiny
\begin{tabular}{|c||c|c|c|} \hline
$z^*_k$ & $l_{max} = 10$ & $l_{max} = 20$ & $l_{max} = 30$ \\ \hline\hline
$z^*_0$ & $3.8716018937757\cdot 10^{-1}$ & $3.9015294442769\cdot 10^{-1}$ & $3.9015294689101\cdot 10^{-1}$ \\\hline
$z^*_1$ & $2.7472709654097\cdot 10^{-1}$ & $2.7468216314596\cdot 10^{-1}$ & $2.7468216310166\cdot 10^{-1}$ \\ \hline
$z^*_2$ & $2.0416445882152\cdot 10^{-1}$ & $2.0327597434157\cdot 10^{-1}$ & $2.0327597360790\cdot 10^{-1}$ \\ \hline
$z^*_3$ & $1.4092248806621\cdot 10^{-1}$ & $1.4013502479152\cdot 10^{-1}$ & $1.4013502414472\cdot 10^{-1}$ \\ \hline
$z^*_4$ & $8.9625774854293\cdot 10^{-2}$ & $8.9257511848321\cdot 10^{-2}$ & $8.9257511547236\cdot 10^{-2}$ \\ \hline
$z^*_5$ & $5.2733434885004\cdot 10^{-2}$ & $5.2748796970937\cdot 10^{-2}$ & $5.2748796984464\cdot 10^{-2}$ \\ \hline
$z^*_6$ & $2.8856392240700\cdot 10^{-2}$ & $2.9099303116770\cdot 10^{-2}$ & $2.9099303316272\cdot 10^{-2}$ \\ \hline
$z^*_7$ & $1.4747091527680\cdot 10^{-2}$ & $1.5072774239603\cdot 10^{-2}$ & $1.5072774508427\cdot 10^{-2}$ \\ \hline
$z^*_8$ & $7.0528086140537\cdot 10^{-3}$ & $7.3689239174505\cdot 10^{-3}$ & $7.3689241862291\cdot 10^{-3}$ \\ \hline
$z^*_9$ & $3.1541868355104\cdot 10^{-3}$ & $3.4157386267627\cdot 10^{-3}$ & $3.4157388689480\cdot 10^{-3}$ \\ \hline
$z^*_9$ & $1.3139603409337\cdot 10^{-3}$ & $1.5071112742866\cdot 10^{-3}$ & $1.5071114902056\cdot 10^{-3}$ \\ \hline
\vdots & - & \vdots & \vdots  \\ \hline
$z^*_{20}$ & -                & $5.0275646988651\cdot 10^{-8}$ & $5.0374367835756\cdot 10^{-8}$ \\ \hline
\vdots & - & - &\vdots \\ \hline
$z^*_{30}$ & -                & -                   & $1.0553957852574\cdot 10^{-13}$ \\ \hline
\end{tabular}
}
}
\caption{Effects of the truncation for $N=3$}
\label{N3table}
\end{table}

Table \ref{N3table} shows the first 10 coefficients of the fixed point $z^*_k$ for
$N=3, d=3$. If we go to 
$l_{max} = 40$ the first 10 coefficients are the same as for $l_{max} = 30$.  For
$l_{max} = 40$ we found $z^*_{40} = 3.2209843597592\cdot 10^{-20}$, $z^*_{30} =
1.0554235840436\cdot 10^{-13}$ and
$z^*_{20} = 5.0374367851155\cdot 10^{-8}$. So we can use $l_{max} = 30$ for $N=3$.

\begin{table}[htp]
\centerline{
{\tiny
\begin{tabular}{|c||c|c|c|} \hline
$z^*_k$ & $l_{max} = 40$ & $l_{max} = 50$ & $l_{max} = 60$  \\ \hline\hline
$z^*_0$ & $6.7135699167269\cdot 10^{-4}$ & $6.7137831100317\cdot 10^{-4}$ & $6.7137831236764\cdot 10^{-4}$ \\ \hline
$z^*_1$ & $5.4722582473628\cdot 10^{-4}$ & $5.4724091850071\cdot 10^{-4}$ & $5.4724091946673\cdot 10^{-4}$ \\ \hline
$z^*_2$ & $5.2202412254520\cdot 10^{-4}$ & $5.2203651652976\cdot 10^{-4}$ & $5.2203651732299\cdot 10^{-4}$ \\ \hline
$z^*_3$ & $5.0324322565809\cdot 10^{-4}$ & $5.0325339440537\cdot 10^{-4}$ & $5.0325339505618\cdot 10^{-4}$ \\ \hline
$z^*_4$ & $4.7793545596989\cdot 10^{-4}$ & $4.7794355811177\cdot 10^{-4}$ & $4.7794355863032\cdot 10^{-4}$ \\ \hline
$z^*_5$ & $4.4366461453368\cdot 10^{-4}$ & $4.4367080895417\cdot 10^{-4}$ & $4.4367080935062\cdot 10^{-4}$ \\ \hline
$z^*_6$ & $4.0145889929024\cdot 10^{-4}$ & $4.0146340398981\cdot 10^{-4}$ & $4.0146340427811\cdot 10^{-4}$ \\ \hline
$z^*_7$ & $3.5381684226392\cdot 10^{-4}$ & $3.5381992679245\cdot 10^{-4}$ & $3.5381992698987\cdot 10^{-4}$ \\ \hline
$z^*_8$ & $3.0372937145831\cdot 10^{-4}$ & $3.0373132889402\cdot 10^{-4}$ & $3.0373132901930\cdot 10^{-4}$ \\ \hline
$z^*_9$ & $2.5407316573042\cdot 10^{-4}$ & $2.5407428246010\cdot 10^{-4}$ & $2.5407428253157\cdot 10^{-4}$ \\ \hline
$z^*_9$ & $2.0723798082894\cdot 10^{-4}$ & $2.0723851264509\cdot 10^{-4}$ & $2.0723851267913\cdot 10^{-4}$ \\ \hline
\vdots & \vdots &  \vdots & \vdots  \\ \hline
$z^*_{36}$ & $1.1315164646491\cdot 10^{-9}$ & $1.1338793621549\cdot 10^{-9}$ & $1.1338796345545\cdot 10^{-9}$ \\ \hline
$z^*_{37}$ & $5.7800321411117\cdot 10^{-10}$ & $5.7989994920249\cdot 10^{-10}$ & $5.7990021959982\cdot 10^{-10}$ \\ \hline
$z^*_{38}$ & $2.9151584691444\cdot 10^{-10}$ & $2.9299955707748\cdot 10^{-10}$ & $5.7990021959982\cdot 10^{-10}$ \\ \hline
$z^*_{39}$ & $1.4515862129964\cdot 10^{-10}$ & $1.4628783782344\cdot 10^{-10}$ & $1.4628809576271\cdot 10^{-10}$ \\ \hline
$z^*_{40}$ & $7.1353999581577\cdot 10^{-11}$ & $7.2189285198076\cdot 10^{-11}$ & $7.2189530200557\cdot 10^{-11}$ \\ \hline
\vdots & - & \vdots &  \vdots  \\ \hline
$z^*_{50}$ & -                   & $3.3911297113494\cdot 10^{-14}$ & $3.3943990779867\cdot 10^{-14}$ \\ \hline
\vdots & - & - &  \vdots  \\ \hline
$z^*_{60}$ & -                   & -                   & $6.0217088064527\cdot 10^{-18}$ \\ \hline
\end{tabular}
}
}
\caption{Effects of the truncation for $N=20$}
\label{N20table}
\end{table}

$N=20$ was the greatest value of $N$ used during numerical calculations. See table 
\ref{N20table} for a selection of coefficients of the fixed point. One can observe the
increasing difficulty of finding the fixed point because the first 10 coefficients have
the same order of magnitude and they are quite small. One has to go to $l_{max} = 60$
before the first 10 coefficients are constant under further increasing of $l_{max}$. Note
further that the coefficients of the fixed point do not decrease as
fast as for lower values of $N$.

By increasing $l_{max}$ to $70$ the first 10 coefficients are identical to those for
$l_{max} = 60$. We found for
$l_{max} = 70$: $z^*_{60} = 6.0220916546711\cdot 10^{-18}$, $z^*_{50} =
3.3943995424453\cdot 10^{-14}$ and
$z^*_{40} = 7.2189530204572\cdot 10^{-11}$.

\section{Numerical results: the exponent $\nu$}

\begin{table}[htp]
\centerline{
{\footnotesize
  \begin{tabular}{|r||c|c||r|c|c|} \hline
   $d$   & $N = -1$ & $N = 0$ & $d$ & $N = -1$ & $N = 0$ \\  \hline\hline
   $0$   & -            & -            & $2$   & $0.572496$ & $0.766551$  \\ \hline
   $0.1$ & -            & -            & $2.1$ & $0.570806$ & $0.742861$  \\ \hline
   $0.2$ & -            & -            & $2.2$ & $0.568709$ & $0.720999$  \\ \hline
   $0.3$ & -            & $3.357170$ & $2.3$ & $0.566240$ & $0.700752$  \\ \hline
   $0.4$ & $0.528092$ & $2.557642$ & $2.4$ & $0.563437$ & $0.681937$  \\ \hline
   $0.5$ & $0.534783$ & $2.094339$ & $2.5$ & $0.560333$ & $0.664404$  \\ \hline
   $0.6$ & $0.541065$ & $1.794838$ & $2.6$ & $0.556966$ & $0.648025$  \\ \hline
   $0.7$ & $0.546882$ & $1.585792$ & $2.7$ & $0.553369$ & $0.632691$  \\ \hline
   $0.8$ & $0.552197$ & $1.431311$ & $2.8$ & $0.549577$ & $0.618308$  \\ \hline
   $0.9$ & $0.556982$ & $1.312021$ & $2.9$ & $0.545622$ & $0.604796$  \\ \hline
   $1$   & $0.561217$ & $1.216665$ & $3$   & $0.541536$ & $0.592086$  \\ \hline
   $1.1$ & $0.564886$ & $1.138306$ & $3.1$ & $0.537351$ & $0.580118$  \\ \hline
   $1.2$ & $0.567979$ & $1.066439$ & $3.2$ & $0.533094$ & $0.568842$  \\ \hline
   $1.3$ & $0.570489$ & $1.016113$ & $3.3$ & $0.528794$ & $0.558214$  \\ \hline
   $1.4$ & $0.572419$ & $0.967166$ & $3.4$ & $0.524479$ & $0.548199$  \\ \hline
   $1.5$ & $0.573773$ & $0.924112$ & $3.5$ & $0.520177$ & $0.538766$  \\ \hline
   $1.6$ & $0.574564$ & $0.885845$ & $3.6$ & $0.515917$ & $0.529896$  \\ \hline
   $1.7$ & $0.574809$ & $0.851529$ & $3.7$ & $0.511728$ & $0.521574$  \\ \hline
   $1.8$ & $0.574528$ & $0.820526$ & $3.8$ & $0.507645$ & $0.513801$  \\ \hline
   $1.9$ & $0.573748$ & $0.792332$ & $3.9$ & -            & $0.506593$  \\ \hline
  \end{tabular}
}
}
\caption{\small Numerically calculated values of $\nu$}
\label{N0minus1table}
\end{table}

Table \ref{N0minus1table} and table \ref{N110table} show some 
numerically calculated results of the critical exponent $\nu$ for
different values of $N$ and $d$.

\begin{table}[htp]
\centerline{
{\footnotesize 
  \begin{tabular}{|r||c|c|c|c|c|} \hline
   $d$   & $N = 1$ & $N = 2$ & $N = 3$ & $N = 5$ & $N = 10$ \\  \hline\hline
   $2$    & - & - & - & - & - \\ \hline
   $2.1$  & $2.088862$ & $9.193920$ & -            & -            & -   \\ \hline
   $2.2$  & $1.362344$ & $4.045225$ & -            & -            & -   \\ \hline
   $2.3$  & $1.099159$ & $2.288147$ & $2.889098$ & $3.126743$ & -   \\ \hline
   $2.4$  & $0.957038$ & $1.559028$ & $2.030739$ & $2.287148$ & -   \\ \hline
   $2.5$  & $0.865337$ & $1.214032$ & $1.535818$ & $1.782761$ & $1.908031$   \\ \hline
   $2.6$  & $0.799849$ & $1.022244$ & $1.236507$ & $1.449465$ & $1.574435$   \\ \hline
   $2.7$  & $0.749927$ & $0.901226$ & $1.046530$ & $1.217683$ & $1.337054$   \\ \hline
   $2.8$  & $0.710110$ & $0.817609$ & $0.918937$ & $1.051532$ & $1.160436$   \\ \hline
   $2.9$  & $0.677294$ & $0.755940$ & $0.828437$ & $0.929516$ & $1.024914$   \\ \hline
   $3$    & $0.649570$ & $0.708225$ & $0.761140$ & $0.837755$ & $0.918611$   \\ \hline
   $3.1$  & $0.625700$ & $0.669949$ & $0.709101$ & $0.767069$ & $0.833820$   \\ \hline
   $3.2$  & $0.604841$ & $0.638382$ & $0.667563$ & $0.711338$ & $07652475$   \\ \hline
   $3.3$  & $0.586400$ & $0.611780$ & $0.633545$ & $0.666445$ & $0.699111$   \\ \hline
   $3.4$  & $0.569946$ & $0.588980$ & $0.605103$ & $0.629584$ & $0.662617$   \\ \hline
   $3.5$  & $0.555163$ & $0.569176$ & $0.580926$ & $0.598813$ & $0.623711$   \\ \hline
   $3.6$  & $0.541815$ & $0.551801$ & $0.560104$ & $0.572765$ & -   \\ \hline
   $3.7$  & $0.529725$ & $0.536443$ & - & $0.546388$ & -   \\ \hline
   $3.8$  & $0.518769$ & $0.522810$ & $0.526130$ & $0.531208$ & -   \\ \hline
   $3.9$  & $0.508867$ & $0.510699$ & $0.512199$ & $0.514498$ & -   \\ \hline
  \end{tabular}
}
}
\caption{\small Numerically calculated values of $\nu$}
\label{N110table}
\end{table}

\section{Some results of the $\epsilon$ expansion}
\label{epsilonappendix}

In this appendix we present some of the results of the $\epsilon$ expansion calculated using
{\sf Maple} 

\subsection{$\epsilon$ expansion of the coefficients $z_m$ up to 2. order}

$T = \sqrt{2}$ und $R = \ln 2$
{\footnotesize
\begin{eqnarray*}
z_0^{(0)} 
  & = & 
    1
  \\
z_0^{(1)}
  & = & 
    0
  \\
z_0^{(2)}
  & = & 
  \frac {-N R^2( N + 2)}{32(N+8)^2}
   \\
z_1^{(0)}
  & = & 
   0
 \\
z_1^{(1)}
  & = & 
   0
  \\
z_1^{(2)}
  & = & 
   \frac {R^2 (N + 2)}{8(T-2)(N+8)^2}
   \\
z_2^{(0)}
  & = & 
   0
 \\
z_2^{(1)}
  & = & 
   -\frac{R}{16(N+8)}
  \\
z_2^{(2)}
  & = & R\Big(-8(N+8)^2 + 4(- N^2 + 44 N + 200) R + 6(N+8)^2 T + \\
& &  (-92 N - 424 + 3 N^2)T R \Big)
      \left/ {\vrule height0.37em width0em depth0.37em} \right. 
      \Big( (N+8)^3(-384 T +  512) \Big)
   \\
z_3^{(0)}
  & = & 
   0
 \\
z_3^{(1)}
  & = & 
   0
  \\
z_3^{(2)}
  & = & 
   \frac{R^2}{32(N+8)^2 (T - 1)}
   \\
z_4^{(0)}
  & = & 
   0
 \\
z_4^{(1)}
  & = & 
   0
  \\
z_4^{(2)}
  & = & 
   \frac{R^2}{512(N+8)^2}
   \\
\end{eqnarray*}
}

\subsection{$\epsilon$ expansion of the exponent $\nu$}
{\footnotesize
\begin{eqnarray*}
\nu^{(0)} & = & \frac 1 2 \\
\nu^{(1)} & = & \frac 1 4 \frac{N+2}{N+8} \\
\nu^{(2)} & = & -\frac{(N+2)} {16(N+8)^3(4-3 T)} \cdot \\
& &
 \Big( - 8(N + 8)(N + 2)
 + 12(7 N + 20) R
   \\
   & &
\;\;\;\; \Big. 6(N + 8)(N + 2) T
 - 7(7 N + 20) R T \Big)
  \\
\nu^{(3)} & = & -\frac{(N+2)} {32(N+8)^5(4-3T)^2} \cdot \\
& &
 \Big(- 68(N + 8)^2 (N + 2)^2 + 45(N + 8)(7 N + 20)(5 N + 16) R \Big.
\\
& &
\;\;\;\; (681 N^3 - 440 N^2 - 2648 N + 42016) R^2 + 48 (N + 8)^2 (N + 2)^2 T
\\
& &
\;\;\;\; \Big.- 32(N + 8)(7 N + 20)(5 N + 16) R T
\\
& &
\;\;\;\; 
\Big.- 6 (79 N^3 - 60 N^2 - 432 N + 4544) R^2 T \Big)
\\
\end{eqnarray*}
}

The zeroth and first order of the $\epsilon$ expansion are the same for the hierarchical and the
full model. For the full model we have
{\footnotesize
$$
  \nu^{(2)}_f = \frac{N+2}{8(N+8)^3}(N^2+23 N+60)
$$}
and
\begin{eqnarray*}
  \nu^{(3)}_f & = & \frac{N+2}{32(N+8)^5}(2 N^4 + 89 N^3 + 1412 N^2 + 5904 N + \\
& & \;\;\; 8640 -
192(5 N + 22)(N+8)t)
\end{eqnarray*}
with $t \approx 0.60103$ \cite{wilsonkogut}.
Note that the prefactors 
$\frac{N+2}{(N+8)^3}$ and
$\frac{N+2}{(N+8)^5}$ are the same in both expansions and that the limit 
$N\rightarrow\infty$ is the same for both models and equals the coefficients of the expansion of 
$\frac{1}{d-2} = \frac{1}{2} \frac{1}{1-\epsilon/2}$. ($\nu^{(1)}\rightarrow \frac
1 4, \nu^{(2)} \rightarrow \frac 1 8$ and $\nu^{(3)}
\rightarrow \frac 1 {16}$)

\begin{table}[htp]
{\tiny
  \begin{tabular}{|r|ccccccccc|} \hline
$N$  & $3.8$ & $3.7$ & $3.6$ & $3.5$ & $3.4$ & $3.3$& $3.2$& $3.1$ & $3.0$\\  \hline\hline
$-2$ & 0.50000 & 0.50000 & 0.50000 & 0.50000 & 0.50000 & 0.50000 & 0.50000 & 0.49996 & 0.49994 \\ 
$-1$ & 0.50764 & 0.51172 & 0.51586 & 0.51995 & 0.52371 & 0.52744 & 0.53299 & 0.53266 &      -  \\ 
$0$  & 0.51380 & 0.52157 & 0.52988 & 0.53873 & 0.54811 & 0.55761 & 0.56841 & 0.57946 & 0.59040 \\
$1$  & 0.51877 & 0.52973 & 0.54183 & 0.55525 & 0.56972 & 0.58789 & 0.60133 & 0.62597 &      -  \\
$2$  & 0.52281 & 0.53644 & 0.55181 & 0.56920 & 0.58817 & 0.61332 & 0.63773 & 0.67734 &      -  \\
$3$  & 0.52613 & 0.54199 & 0.56010 & 0.58093 & 0.60520 & 0.63391 &    -    &  -      &      -  \\
$4$  & 0.52889 & 0.54660 & 0.56699 & 0.59064 & 0.61836 & 0.65126 & 0.69084 & 0.73922 & 0.79939 \\
$5$  & 0.53121 & 0.55046 & 0.57274 & 0.59872 & 0.62928 & 0.66562 & 0.70931 & 0.76246 & 0.82801 \\
$6$  & 0.53318 & 0.55372 & 0.57757 & 0.60546 & 0.63834 & 0.67747 & 0.72449 & 0.78161 & 0.85185 \\
$7$  & 0.53487 & 0.55650 & 0.58167 & 0.61113 & 0.64589 & 0.68727 & 0.73696 & 0.79724 & 0.87121 \\
$8$  & 0.53633 & 0.55890 & 0.58517 & 0.61593 & 0.65224 & 0.69543 & 0.74724 & 0.81000 & 0.88690 \\
$9$  & 0.53761 & 0.56098 & 0.58818 & 0.62004 & 0.65761 & 0.70227 & 0.75578 & 0.82050 & 0.89964 \\
$10$ & 0.53873 & 0.56279 & 0.59079 & 0.62357 & 0.66220 & 0.70806 & 0.76293 & 0.82919 & 0.91008 \\
$11$ & 0.53973 & 0.56439 & 0.59308 & 0.62664 & 0.66615 & 0.71300 & 0.76897 & 0.83646 & 0.91868 \\
$12$ & 0.54061 & 0.56581 & 0.59510 & 0.62933 & 0.66958 & 0.71725 & 0.77412 & 0.84258 & 0.92585 \\
$13$ & 0.54141 & 0.56707 & 0.59688 & 0.63169 & 0.67258 & 0.72093 & 0.77854 & 0.84778 & 0.93187 \\ 
$14$ & 0.54212 & 0.56821 & 0.59847 & 0.63378 & 0.67522 & 0.72415 & 0.78237 & 0.85225 & 0.93697 \\
$15$ & 0.54277 & 0.56923 & 0.59990 & 0.63565 & 0.67755 & 0.72698 & 0.78572 & 0.85611 & 0.94133 \\ 
$16$ & 0.54336 & 0.57015 & 0.60119 & 0.63732 & 0.67963 & 0.72949 & 0.78866 & 0.85948 & 0.94510 \\
$17$ & 0.54390 & 0.57099 & 0.60235 & 0.63883 & 0.68150 & 0.73172 & 0.79126 & 0.86243 & 0.94838 \\
$18$ & 0.54440 & 0.57176 & 0.60341 & 0.64019 & 0.68318 & 0.73372 & 0.79357 & 0.86504 & 0.95125 \\
$19$ & 0.54485 & 0.57246 & 0.60438 & 0.64143 & 0.68469 & 0.73551 & 0.79564 & 0.86737 & 0.95379 \\
$20$ & 0.54527 & 0.57311 & 0.60526 & 0.64256 & 0.68607 & 0.73714 & 0.79751 & 0.86945 & 0.95605 \\
\hline\hline
$\frac{1}{d-2}$ 
     & 0.55556 & 0.58823 & 0.62500 & 0.66667 & 0.71429 & 0.76923 & 0.83333 & 0.90909 & 1.00000 \\ 
\hline
  \end{tabular}
}
\caption{$\nu$: Diagonal Borel-Pad\'{e} summation of the $\epsilon$-expansion
up to 6. order.}
\label{epsbptable}
\end{table}

\section{Figures}

\clearpage

\renewcommand{\textfraction}{0.0}
\renewcommand{\topfraction}{1.0}
\renewcommand{\bottomfraction}{1.0}

\clearpage


\end{appendix}


\begin{thebibliography}{10}

\bibitem{wegnerhoughtonrg}
K.~I. Aoki, K.~I. Morikawa, W.~Souma, J.~I. Sumi, and H.~Terao.
\newblock The effectiveness of the local potential approximation in the
  {Wegner-Houghton} renormalization group.
\newblock {\em hep-ph/9612458,
Prog. Theor. Phys.}, 95:409--420, 1996.

\bibitem{RGApproachToSurfacesCassandro}
M.~Cassandro and P.K. Mitter.
\newblock Renormalization group approach to interacting crumpled surfaces:
  {The} hierarchical recursion.
\newblock {\em Nucl. Phys. B}, 422:634--676, 1994.

\bibitem{comellastravesset}
J.~Comellas and A.~Travesset.
\newblock {$O(N)$} models within the local potential approximation.
\newblock {\em hep-th/9701028, 
Nucl. Phys. B}, 498:539-564, 1997.

\bibitem{dipl}
J.~G\"ottker-Schnetmann.
\newblock {Analytische} {und} {numerische} {Untersuchungen} {hierarchischer}
  {Renormierungsgruppenfixpunkte} {am} {Beispiel} {$O(N)$}-{invarianter}
  {Modelle}.
\newblock Diploma thesis, Westf\"alische Wilhelms-Universit\"at M\"unster, 1
  1996.

\bibitem{guidazinnjustin}
R.~Guida and J~Zinn-Justin.
\newblock Critical exponents of the $n$-vector model.
\newblock {\em cond-mat/9803240},
  1998.

\bibitem{kochwittwer}
H.~Koch and P.~Wittwer.
\newblock On the renormalization group transformation for scalar hierarchical
  models.
\newblock {\em Comm. Math. Phys.}, 138:537--568, 1991.

\bibitem{pipowi}
K.~Pinn, A.~Pordt, and C.~Wieczerkowski.
\newblock Algebraic computation of hierarchical renormalization group fixed
  points and their $\epsilon$-expansions.
\newblock {\em
hep-lat/9402020, J.
  Stat. Phys}, 77:977--1005, 1994.

\bibitem{pirewi}
K.~Pinn, M.~Rehwald, and C.~Wieczerkowski.
\newblock On the stability of the {O(N)}-invariant and the cubic-invariant
  3-dimensional {N}-component renormalization group fixed points in the
  hierarchical approximation.
\newblock {\em
  cond-mat/9805193,
  J. Stat. Phys.}, 95:1--22, 1999.

\bibitem{nonassPordtWiec}
A.~Pordt and C.~Wieczerkowski.
\newblock Nonassociative algebras and nonperturbative field theory for
  hierarchical models.
\newblock
  {\em hep-lat/9406005},
  Preprint MS-TPI-94-4, Westf\"alische Wilhelms-Universit\"at M\"unster, 4
  1994.

\bibitem{reiszhightemperatureON}
T.~Reisz.
\newblock High temperature critical {$O(N)$} field models by {LCE} series.
\newblock {\em Physics Letters B}, 360:77--82, 1995.

\bibitem{wilsonkogut}
K.~G. Wilson and J.~Kogut.
\newblock The renormalization group and the $\epsilon$-expansion.
\newblock {\em Physics Letters C}, 2:75--200, 1974.

\end{thebibliography}
\end{document}